\documentclass[aps,prb,superscriptaddress,showpacs,twocolumn,longbibliography]{revtex4-1}
\usepackage{amsmath, bm}
\usepackage{amssymb}
\usepackage{graphicx}
\usepackage{hyperref}
\usepackage{color}
\usepackage{hhline}
\usepackage{subeqnarray}

\newcommand{\DM}{Dzyaloshinskii-Moriya }

\begin{document}
\title{Emergent Potts order in the kagom\'e \texorpdfstring{$J_1-J_3$}{J1--J3} Heisenberg model}

\author{Vincent Grison}
\affiliation{Sorbonne Universit\'e, CNRS, Laboratoire de Physique Th\'eorique de la Mati\`ere Condens\'ee (LPTMC), F-75005 Paris, France}

\author{Pascal Viot}
\email{viot@lptmc.jussieu.fr}
\affiliation{Sorbonne Universit\'e, CNRS, Laboratoire de Physique Th\'eorique de la Mati\`ere Condens\'ee (LPTMC), F-75005 Paris, France}

\author{Bernard Bernu}
\email{bernu@lptmc.jussieu.fr}
\affiliation{Sorbonne Universit\'e, CNRS, Laboratoire de Physique Th\'eorique de la Mati\`ere Condens\'ee (LPTMC), F-75005 Paris, France}

\author{Laura Messio}
\email{messio@lptmc.jussieu.fr}
\affiliation{Sorbonne Universit\'e, CNRS, Laboratoire de Physique Th\'eorique de la Mati\`ere Condens\'ee (LPTMC), F-75005 Paris, France}
\affiliation{Institut Universitaire de France (IUF), 1 rue  Descartes, F-75005 Paris,  France}

\begin{abstract}
Motivated by the physical properties of Vesignieite BaCu$_3$V$_2$O$_8$(OH)$_2$, we study the $J_1-J_3$ Heisenberg model on the kagom\'e lattice, that is proposed to describe this compound for $J_1<0$ and $J_3\gg|J_1|$. 
The nature of the classical ground state and the possible phase transitions are investigated through analytical calculations and parallel tempering Monte Carlo simulations. 
For $J_1<0$ and $J_3>\frac{1+\sqrt{5}}4|J_1|$, the ground states are not all related by an Hamiltonian symmetry. 
Order appears at low temperature via the order by disorder mechanism, favoring colinear configurations and leading to an emergent $q=4$ Potts parameter. 
This gives rise to a finite temperature phase transition. 
Effect of quantum fluctuations are studied through linear spin wave approximation and high temperature expansions of the $S=1/2$ model.  
For $J_3$ between $\frac14|J_1|$ and $\frac{1+\sqrt{5}}4|J_1|$, the ground state goes through a succession of semi-spiral states, possibly giving rise to multiple phase transitions at low temperatures. 
\end{abstract}

\date{\today}

\maketitle

\section{Introduction}

The existence of competing interactions in a magnetic spin lattice model leads to the inability to satisfy all pair interactions simultaneously. 
The system is said to be \textit{frustrated}. 
While its effects in a classical spin model can be important, they are enforced for quantum spin models, where they may induce spin liquid ground states\cite{Review_SL}. 
These phases break none of the Hamiltonian symmetries and as a consequence, show no magnetic long range order. 
Thus, it is interesting to pick up classical models where frustration has the largest effects, 
in view to detect quantum models hosting highly disordered phases. 

Such spin models on the bidimensional kagom\'e lattice have a long history, both from theoretical and experimental point of view. 
The most studied model is definitely the first neighbor antiferromagnet, realized in Herbertsmithite\cite{Vries2009}, even if impurities and other interactions keep this compound away from its idealization.  
In the search of the perfect chemical realization of this specific model, many other kagom\'e compounds were proposed, such as 
Kapellasite\cite{Kapellasite_cuboc2}, Volborthite\cite{volbVSherb},
Haydeite\cite{Colman2010,Colman2011}, 
Ba-Vesignieite\cite{JPSJ.78.033701,c2jm32250a,PhysRevB.83.180407, Ishikawa2017}, 
Sr-Vesignieite\cite{PhysRevB.101.054425,PhysRevB.101.054425}... although they were finally described by different interactions. 
Here we shall restrict our attention to the model  supposed to describe the Ba-Vesignieite compound\cite{PhysRevLett.121.107203}, with small first neighbor ferromagnetic and large third neighbor antiferromagnetic interactions. 

In the Vesignieite BaCu$_3$V$_2$O$_8$(OH)$_2$ compound, magnetic Cu atoms form decoupled and perfect bidimensional kagom\'e layers of $S=1/2$ spins. 
Its Curie-Weiss temperature is around $-77K$\cite{JPSJ.78.033701}, indicating an antiferromagnetic dominant coupling that was first proposed to be first neighbor\cite{JPSJ.78.033701}. 
Moreover, specific heat, magnetic susceptibility and powder neutron diffraction measurements on Vesignieite were supporting the spin liquid ground state hypothesis\cite{JPSJ.78.033701, PhysRevB.83.180416}, even if more and more indications of a phase transition around 9K appeared with time\cite{PhysRevB.83.180416, PhysRevB.84.180401}. 
This transition, probably related to a small interlayer coupling, is now clearly identified in crystalline samples\cite{c2jm32250a}. 
Finally, neutron diffraction results on crystals\cite{PhysRevB.83.180416} indicated that the short range spin correlations were uncompatible with antiferromagnetic first neighbor interactions ($J_1$ in Fig.~\ref{fig:neighbor_kagome}), but coherent with a dominant third neighbor interaction $J_3$. 
These unusual interactions in Ba-Vesignieite are our main motivation to explore this kagome model. 
To the best of our knowledge, the $J_1-J_3$ Heisenberg model on the kagom\'e lattice\cite{PhysRevLett.121.107203} has still not been studied for large $J_3$.  

In Heisenberg models, the interaction between two tridimensional unit spins on sites $i$ and $j$ is given by $J_{i,j}\,\mathbf S_i\cdot \mathbf S_j$. 
$J_{i,j}$ is the coupling constant, either positive for antiferromagnetic interactions, or negative for ferromagnetic ones. 
Unfrustated classical Heisenberg models have colinear ground states (i.e. all the spins are oriented along a unique line, with only two possible directions). 
It is notably the case for ferromagnetic models, of for antiferromagnetic ones on bipartite lattices, where sites can be labelled $A$ or $B$ in such a way that only different types of sites interact. 
Frustration can induce non-colinear magnetic orders, as on the triangular lattice with antiferromagnetic interactions: three sublattices $A$, $B$ and $C$ host spins directions $\mathbf S_A$, $\mathbf S_B$ and $\mathbf S_C$ each at an angle of 120$^\circ$ from the others. 
In this case, spins are no more colinear but remain coplanar. 
More rarely, non-coplanar spin states are obtained in Heisenberg models\cite{KagomeDomenge, cuboc1, Kapellasite_cuboc2, PhysRevB.83.184401, Sklan2013}, with possibly large unit cells. 
Twelve-site unit cells, with spins pointing towards the corners of a cuboctahedron are for example obtained on the kagom\'e lattice for interactions up to third neighbors\cite{KagomeDomenge, cuboc1, Kapellasite_cuboc2}. 

The Mermin-Wagner theorem states that no continuous symmetry of a Hamiltonian can be broken at finite (non-zero) temperature in two dimensions\cite{MerminWagner, Mermin1967, KleinLandauShucker}. 
Yet, other types of finite temperature phase transitions exist, relating phases with or without symmetry breaking\cite{KosterlitzThouless, Hilhorst}, associated with topological defects for instance. 
When a Hamiltonian symmetry is broken, the Mermin-Wagner theorem implies that it is a discrete one. 
In Heisenberg models, global spin rotations form a continuous symmetry group, thus the broken symmetry is different: it can be a lattice symmetry\cite{Zhitomirsky1996}, or the time reversal symmetry\cite{MessioDomenge, Triedres}. 
In most cases, a phase transition can be inferred from the analysis of the ground state manifold: several connected components generally correspond to a broken discrete symmetry. 
For example, if the spins are non coplanar, the ground state manifold is isomorphic to $O(3)$, which has two connected components $\pm SO(3)$.
An emergent Ising parameter $\pm1$ (chirality) can be defined, indicating in which connected component the spin state is. 
The time reversal symmetry ($\mathbf S_i\to-\mathbf S_i$) is broken in the ground state but is restored at finite temperature via a phase transition\cite{KagomeDomenge, MessioDomenge, Triedres}. 

In most cases, all the ground states are equivalent, in the sense that they are related by a symmetry of the Hamiltonian. 
For example, two ground states of the triangular antiferromagnetic lattice each have three different spin orientations on their sublattices: $\mathbf S_A$, $\mathbf S_B$, $\mathbf S_C$ and $\mathbf S'_A$, $\mathbf S'_B$, $\mathbf S'_C$. 
But there exists a three dimensional rotation $R$ such that
$$\forall \alpha \in \{A,B,C\},\, \mathbf S'_\alpha = R\,\mathbf S_\alpha.$$ 
$R$ is an Hamiltonian symmetry: for any spin configuration, the $R$-transformed one has the same energy. 
When the symmetries of the Hamiltonian fail to make all of the ground states equivalent, we speak of accidental degeneracy. Different ground states then have different properties, including different density of low energy excitations. 
This implies that, at low temperatures, some of the ground states are selected by the \textit{order by disorder} mechanism. 
A connected manifold of ground states can thus be reduced to disconnected components at infinitesimal temperatures, possibly giving rise to phase transitions with an emergent discrete order parameter. 
It is precisely what occurs in some part of the phase diagram of the $J_1-J_3$ kagom\'e Heisenberg model, and is the subject of this article. 

The paper is organized as follows. 
In Sec.~\ref{sec:model_and_GS}, we present the model and its classical ground states. 
In  Sec.~\ref{sec:order_by_disorder}, the known examples of order by disorder-induced phase transitions are detailed.
Sec.~\ref{sec:octa_phase} is first devoted to a discussion of the octahedral phase found in the range of parameters corresponding to the Ba-Vesignieite compound (Sec.~\ref{sec:order_by_disorder}), leading in a second part to the definition of an appropriate order parameter (Sec.~\ref{sec:order_parameter}), opening the possibility of a related phase transition.
The finite temperature phase diagram of the classical model is explored using parallel tempering Monte Carlo simulations in Sec.~\ref{sec:MC} and thermal linear spin wave calculations in Sec.~\ref{sec:LTSW}.
A phase transition is evidenced through a finite size analysis, and the critical exponents are numerically evaluated. 
The effects of quantum fluctuations are discussed through a linear spin wave approximation (Sec.~\ref{sec:LQSW}) and high temperature series expansions (Sec.~\ref{sec:HTSE}). 
The relevance of our approach in the case of the $S=1/2$ Ba-Vesignieite compound is discussed.
In conclusion (Sec.~\ref{sec:conclusion}), the nature of the phase transition experimentally observed in Vesignieite is discussed in light of the numerical and analytical results.

\section{The model and its \texorpdfstring{$T=0$}{T=0} classical phase diagram}
\label{sec:model_and_GS}

\begin{figure}
    \centering
        \includegraphics[width=0.4\textwidth]{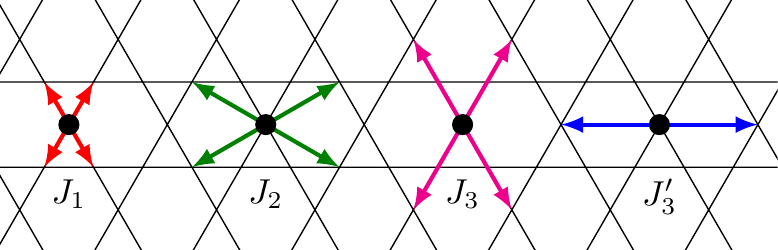}
        \caption{Sketch of first, second and third neighbor interactions on the kagom\'e lattice, $J_1, J_2, J_3$ and $J'_3$ respectively. 
        The third neighbor interaction is split in two contributions: 
        $J_3'$ corresponds to interactions between spins located on two opposite corners of an hexagon, and $J'_3$ between spins located at the same distance, but on corners of two adjacent hexagons.}
        \label{fig:neighbor_kagome}
\end{figure}

The kagom\'e lattice consists of triangles sharing corners, with three sites per unit cell (see Fig.~\ref{fig:neighbor_kagome}). 
On each site $i$, we place a unit vector $\mathbf S_i$ called spin (in the quantum model, they are $S=1/2$ spins). 
For our study, we consider spin interactions between first and third neighbors, with respective strengths $J_1$ and $J_3$ (Fig.~\ref{fig:neighbor_kagome}).
The Hamiltonian of the system reads:
\begin{equation}
\label{eq:HamJ1J3}
\mathcal{H}=J_1\sum_{\langle i,j\rangle} \mathbf S_i \cdot \mathbf S_j
+J_3\sum_{\langle i,j\rangle_3} \mathbf S_i \cdot \mathbf S_j,
\end{equation}
where the sums over $\langle i,j\rangle$ and $\langle i,j\rangle_3$ indicate a sum over all first and third neighbor links of the lattice. 

\begin{figure}
    \begin{center}
        \includegraphics[width=0.4\textwidth]{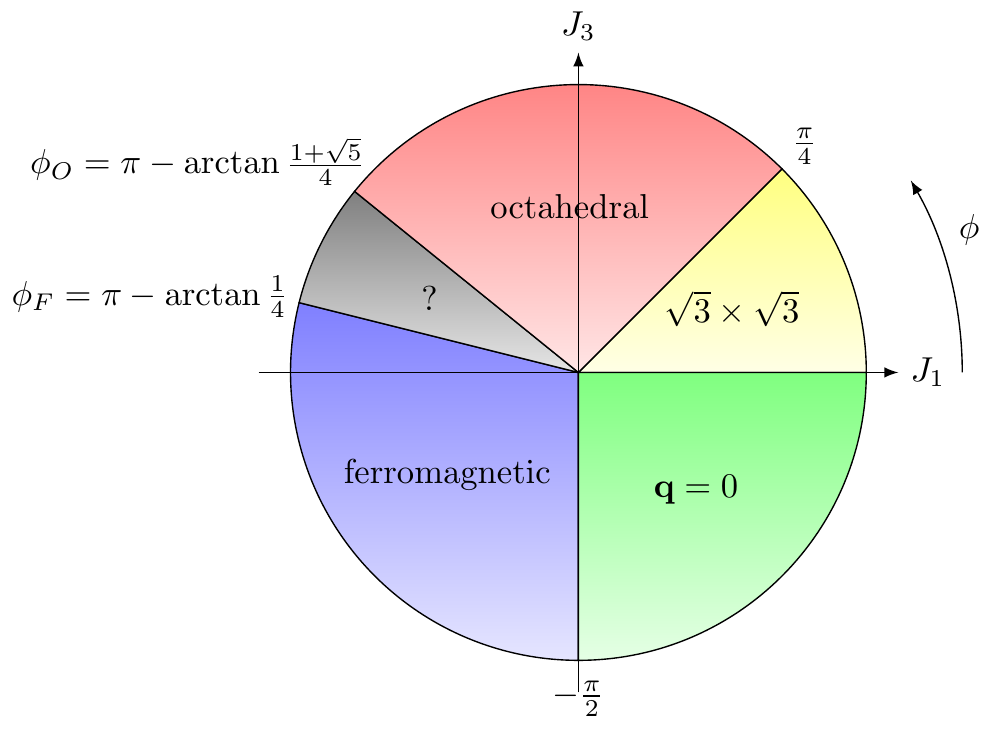}
        \includegraphics[width=0.45\textwidth]{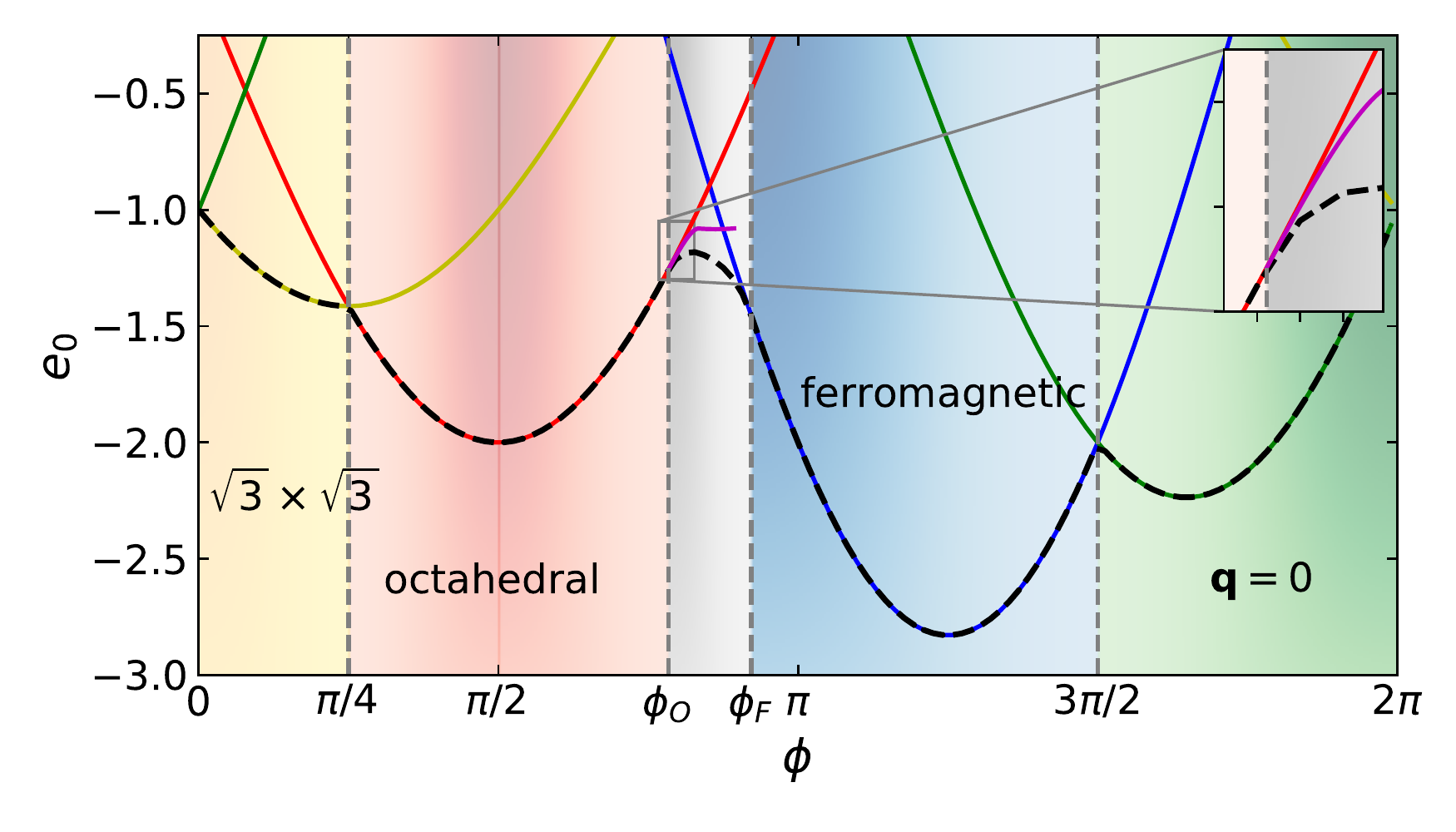}
        \caption{Top: Ground states in the $J_1-J_3$ plane (see Fig.~\ref{fig:neighbor_kagome} for the definition of $J_1$ and $J_3$). 
        The different orders are described in Fig.~\ref{fig:usual_orders_kagome}. 
        Bottom: Energy per site $e_0$ for each state named above, and LT lower bound (dashed). 
        $e_0=2J_1+2J_3$ for the ferromagnetic state (blue line), $-2J_3$ for the octahedral state (red), $-J_1+2J_3$ for the $\mathbf q=0$ state (green) and $-J_1-J_3$ for the $\sqrt3\times\sqrt3$ state (yellow). 
        The lower bound is reached everywhere except in the grey region.
        The magenta curve is the energy of the variational ground state, described in the text and in Fig.~\ref{fig:config_unconventional}.
        }
        \label{fig:camembert} 
    \end{center}
\end{figure}

Let us first investigate the landscape of possible ground states, presented in Fig.~\ref{fig:camembert}. 
We define an energy scale $J=\sqrt{J_1^2+J_3^2}$ and an angle $\phi$ such that $(J_1,J_3)=(J\cos\phi, J\sin\phi)$.

The actual determination of the ground state(s) for given $(J_1,J_3)$ is a tough problem.
No general procedure is known for a classical Hamiltonian such as ours, outside of the case of a quadratic Hamiltonian on a Bravais lattice, that can be handled by the Luttinger-Tizsa (LT) method\cite{Kaplan2007}. 
This method can still be applied in the other cases, but then only gives a lower bound for the ground state energy (see App.~\ref{app:LT}). 
If the energy of a trial state reaches this lower bound, it is then proved to be a ground state. 
Using a group-theoretical approach, a set of spin configurations called \textit{regular magnetic orders} were defined\cite{PhysRevB.83.184401}, that are important trial states. 
In our case, regular magnetic orders are ground states for almost the whole phase diagram, with the exception of a small transition region (grey area of Fig.~\ref{fig:camembert}). 

\begin{figure}
    \begin{center}
        \includegraphics[width=0.45\textwidth]{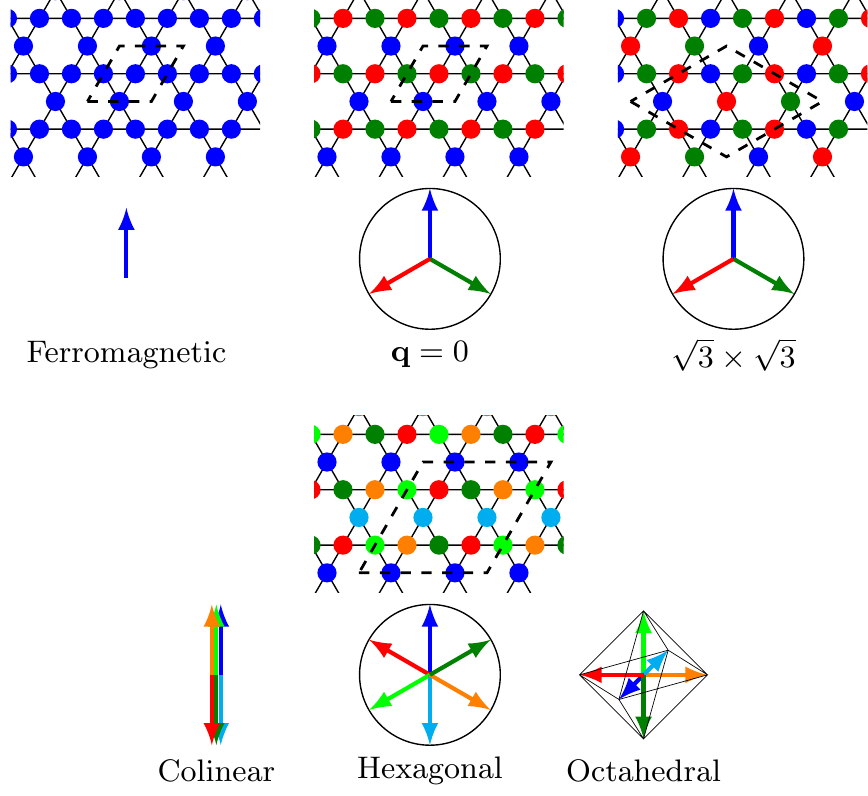}
        \caption{
        Top: Three long-range orders on the kagom\'e lattice, that are ground states in some part of the phase diagram of Fig.~\ref{fig:camembert}. 
        Bottom: Colinear, hexagonal and octahedral states, that belong to the ground state manifold of the octahedral phase of Fig.~\ref{fig:camembert}.
        }
        \label{fig:usual_orders_kagome} 
    \end{center}
\end{figure}

We now describe the the phase diagram of Fig.~\ref{fig:camembert}, whose most phases are described on Fig.~\ref{fig:usual_orders_kagome}. 
When both $J_1$ and $J_3$ are negative, the ground state is obviously a ferromagnetic state, which survives for small positive $J_3$. 
Moving on to an antiferromagnetic coupling $J_1>0$, we encounter the kagom\'e Heisenberg antiferromagnet for $J_3=0$. 
This model is known for its extensive ground-state degeneracy, which is lifted when $J_3$ 
is switched on: 
$J_3 < 0$ aligns spins equivalent under translations of the lattice in different triangles, giving rise to the $\mathbf{q} = 0$ phase, while $J_3 \gtrsim 0$ leads to the $\sqrt{3} \times \sqrt{3}$ order, which survives up to $J_3=J_1$ ($\phi=\pi/4$). 

\begin{figure}
    \begin{center}
        \includegraphics[height=0.2\textwidth]{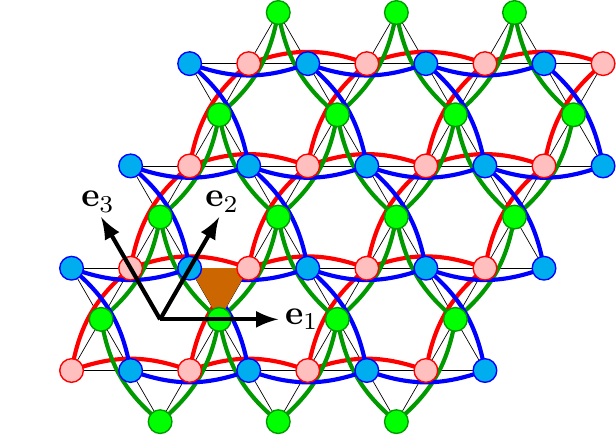}\quad
        \includegraphics[height=0.2\textwidth, trim = 0 -15 0 -15]{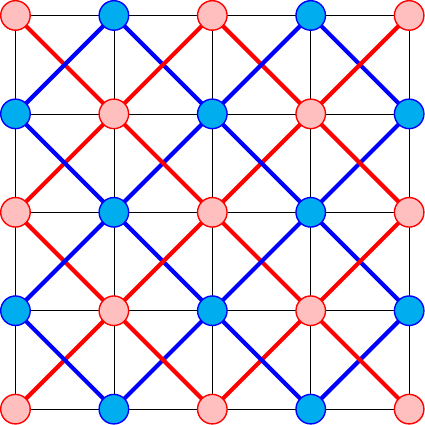}
        \caption{Left: When only $J_3$ interactions are present, the kagom\'e lattice divides into three independent deformed square lattices (with blue, red and green sites and links). 
            When $J_3>0$, an antiferromagnetic $T=0$ spin order sets in on each sublattice, with an arbitrary direction. 
            A small $J_1$ does not lift this degeneracy as it couples for example a red spin with two opposite green spins and two opposite blue spins. 
            The same phenomena occurs on the $J_1-J_2$ square lattice for a strong AF $J_2$ (right).
        }
        \label{fig:Kag_3sublattices} 
    \end{center}
\end{figure}

If $J_1 = 0$, the lattice is decoupled into three square sublattices (Fig.~\ref{fig:Kag_3sublattices}),
each ferromagnetically ($J_3<0$) or antiferromagnetically ($J_3>0$) ordered, in three independent spin directions. 
When $J_3<0$, an infinitesimal (positive or negative) $J_1$ completely lifts the degeneracy towards the ferromagnetic or $\mathbf q=0$ states previously discussed, but this is not the case for $J_3>0$. 
To see why, it is useful to consider a single spin and its nearest neighbors. 
The large value of $J_3$ imposes that each spin is surrounded by pairs of anti-aligned spins, thus \emph{cancelling out} nearest-neighbor energetic contributions as long as each sublattice stays ordered (Fig.~\ref{fig:Kag_3sublattices}).
Thus, a small, arbitrary, $J_1$ does not lift the degeneracy at $T=0$. 
Among the degenerate configurations in this manifold (some of them are illustrated in Fig.~\ref{fig:usual_orders_kagome}), we find a regular order whose spin directions correspond to the vertices of an octahedron, hence the name \textit{octahedral order}\cite{PhysRevB.83.184401}.
At stronger $J_1$, the octahedral state breaks down in favor of other states - $\sqrt{3} \times \sqrt{3}$ for $J_1>0$, and a succession of unconventional states  with eventually several wave vectors for $J_1<0$, before reaching the ferromagnetic sector again. 

\begin{figure}
    \begin{center}
        \includegraphics[width=0.39\textwidth]{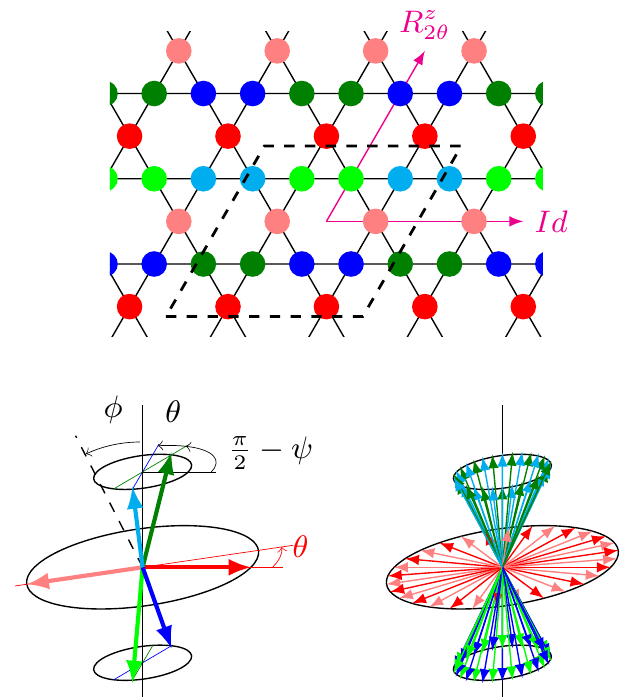}
        \caption{Spin configuration supposed to be the ground state for $\phi$ slightly larger than $\phi_O$, i.e. in the unconventional phase of Fig.~\ref{fig:camembert}. 
        The spins of the dashed unit cell of 12 sites have 6 orientations, as indicated on the bottom left. 
        The parametrization of this state is detailed in App.~\ref{app:phiO}. 
        A translation in the $\mathbf e_1$ direction let the spins invariant, whereas in the $\mathbf e_2$ direction, they are rotated by $2\theta$ around the $z$ axis. 
        Bottom right: orientation of the spins over the full lattice. 
        }
        \label{fig:config_unconventional} 
    \end{center}
\end{figure}

We will now briefly discuss the unconventional ground states of Fig.~\ref{fig:camembert}, even if a detailed description is under the scope of this article. 
In this area of the phase diagram, the LT lower bound of the energy is not reached by any spin configuration and the system has to find a compromise between the different wave vectors to minimize its energy. 
This situation occurs as soon as the wave vector $\mathbf q_{\rm min}$ corresponding to the lowest eigenvalue $\lambda_{\rm min}(\mathbf q)$ becomes different from those of the simple neighboring phases. 
When $\phi$ decreases from $\pi$, we leave the ferromagnetic state at $\phi_F=\pi-\arctan \frac 14$.
The only $\mathbf q_{\rm min}$, previously the zero wave vector, splits into six $\mathbf q_{\rm min}$ staying on lines going from the center of the BZ to its corners. 
When $\phi$ increases, departing from $\pi/2$, we leave the octahedral state at $\phi_O=\pi-\arctan\frac{1+\sqrt{5}}{4}\simeq 0.78\pi$ (proof in App.~\ref{app:phiO}, see also Fig. \ref{fig:LT_J1J_3_kag}).
The three $\mathbf q_{\rm min}$ previously at the middles of the edges of the BZ split into six $\mathbf q_{\rm min}$ staying on lines going from the middles of the edges of the BZ to its center. 
This part of the phase diagram is very rich. 
No method exists to determine the ground states, which usually break several symmetries of the Hamiltonian. 
As an exemple, we describe here the ground state found near $\phi_O$, which is similar to the alternating conic spiral state of\cite{Sklan2013} and whose energy is given in Fig.~\ref{fig:camembert}.  
From numerical simulations (iterative minimization\cite{Sklan2013}), it appears that one of the three sublattices of Fig.~\ref{fig:Kag_3sublattices} develops spin orientations in a plane, say the $xy$ plane, whereas the other two form a cone of axis $z$ and of small angle $\phi$ (see Fig.~\ref{fig:config_unconventional}). 
Note that the orientations of the two last sublattices are exactly the same, translated by a lattice spacing. 
Thus, this state is a spiral state, in the sense given in\cite{PhysRevB.83.184401}, but with an enlarged unit cell of twelve sites, reminiscent of the parent octahedral state.

\section{Ground state selection in the octahedral phase}
\label{sec:octa_phase}

\subsection{Order by disorder}
\label{sec:order_by_disorder}

When $J_1=0$, the three sublattices of Fig.~\ref{fig:Kag_3sublattices} are independent 
and each of them develops its own long range order at zero temperature. 
The ground state is then fully determined by the orientation on three reference sites (say the three sites of a reference unit cell): an element of ${\mathcal S_2}^3$, where ${\mathcal S_2}$ is the unit sphere in three dimensions. 
The effect of a small $J_1$ depends on the sign of $J_3$, as detailed in Sec.~\ref{sec:model_and_GS}. 
For a negative $J_3$, no accidental degeneracy survives to an infinitesimal $J_1$, whatever its sign. 
On the other hand, for positive $J_3$, an infinitesimal $J_1$ has no effect on this degeneracy whatever its sign. 
Note that this accidental degeneracy is not extensive, i.e. does not increase with the lattice size. 
When temperature or quantum fluctuations are switched on, the phenomena of order by disorder occurs, lifting this degeneracy to a subset of ${\mathcal S_2}^3$ - which will be determined below to be $\mathcal S_2\times K_4$, where $K_4$ is the Klein four-group.

Before considering in more detail the kagom\'e $J_1-J_3$ model, let us list some models 
where such (simpler) accidental degeneracies are known. 
Historically, the order by disorder (ObD) phenomenon was described by Villain et al. on a domino model of Ising spins\cite{VillainJ.1980}. 
For a Heisenberg model, the most spectacular and most studied example of ObD is without any doubt the kagom\'e antiferromagnet\cite{Chern2013,Zhitomirsky2002,Schnabel2012,Chernyshev2014,Chernyshev2014,Chernyshev2015,Shender1993,Zhitomirsky2008,Taillefumier2014,Henley2009}, whose degeneracy is extensive, as for the domino model. 
On the kagom\'e lattice, thermal or quantum ObD selects coplanar states, whose number is still extensive, giving rise to possible further ObD effects, such as those occuring in the octupolar order. 

We now focus our attention on other cases of bidimensionnal lattices, which share with the $J_1-J_3$ kagom\'e model a non-extensive accidental degeneracy, with a continuous set of ground states.
This situation is relatively common for Heisenberg Hamiltonians with nearest and next-nearest neighbor interaction. 
A well studied case is  the $J_1-J_2$ Heinsenberg model on a square lattice\cite{Henley1989, PhysRevB.86.184432}, where in the case of strong AF $J_2$, the lattices decouples into 2 sublattices with independent antiferromagnetic orders ($\mathcal S_2^2$ ground state manifold), see Fig.~\ref{fig:Kag_3sublattices}, right. 
Both thermal and quantum fluctuations favor colinear ordering, the ground state manifold being reduced to $\mathcal S_2\times \mathbb Z_2$: the first sublattice has a free orientation ($\mathcal S_2$) and the second one can align its reference spin with the one of the first sublattice, or set it opposite ($\mathbb Z_2$). 
The effective set of ground states is now formed by two disconnected manifolds. 
Depending on the discrete component selected by the system, the $T\to 0^+$ order is an horizontal or vertical columnar state. 
This emergent Ising variable gives rise to a phase transition at finite temperature, compatible with the Mermin-Wagner theorem. 

The Heisenberg models on triangular\cite{Jolicoeur1990} and honeycomb\cite{Fouet2001} lattices also develop ObD favoring colinear states (with a $\mathcal S_2\times \mathbb Z_3$ effective set of ground states) 
for some values of the $J_1-J_2-J_3$ exchanges. 
But contrary to the square lattice, no limit of decoupled lattices allows for an  simple understanding of this phenomenon. 
In the presence of a magnetic field, there are also many expamples of ObD, where colinear configurations 
are stable and lead to magnetization plateaus\cite{Schmidt2017, Gvozdikova2011}. 

\begin{figure}
    \begin{center}
        \includegraphics[width=0.33\textwidth]{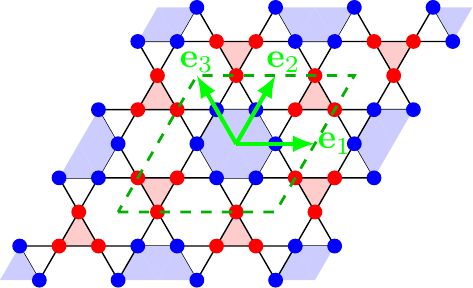}
        \caption{In the octahedral phase of Fig.~\ref{fig:camembert}, order by disorder effect tends to align along a unique direction the spins of the three antiferromagnetic square lattices depicted in Fig.~\ref{fig:Kag_3sublattices}. 
        The resulting colinear spin order has a unit cell of 12 sites (in dashed green) and only two opposite spin orientations (on the blue and red sites).}
        \label{fig:Potts_order} 
    \end{center}
\end{figure}

For the octahedral phase of the $J_1-J_3$ kagom\'e lattice, we can infer from the $J_1-J_2$ square lattice that the three sublattices align their spins colinearly under thermal or quantum fluctuations. 
The ground state manifold thus changes from ${\mathcal S_2}^3$ to $\mathcal S_2\times K_4$: 
the first sublattice has a free orientation ($\mathcal S_2$), the second and third ones can align its reference spin with the one of the first sublattice, or set it opposite (fixing an element of $K_4$) (see Fig.~\ref{fig:Potts_order}). 
The choice of a reference spin for each sublattice is arbitrary, which suggests to use $K_4$ as the symmetry group labelling the different connected components, instead of the isomorphic $\mathbb Z_2^2$, since all symmetries are then explicitely treated on the same footing. 
Note also that the point-group symmetry of the lattice is unchanged - only the translational symmetries are broken. 
$K_4$ is an unusual broken symmetry, but it has already been reported for example in an interacting electron model on the honeycomb lattice\cite{Chern2012}. 

The (effective) ground-state manifold is sometimes abusively called the \textit{order parameter space}. 
We take care here to distinguish them, as an order parameter taking values in another set will be defined in the coming section. 

\subsection{Definition of an order parameter}
\label{sec:order_parameter}

It was envisaged in the preceding section that the ground-state manifold ${\mathcal S_2}^3$ effectively reduces down to $\mathcal S_2\times K_4$ when infinitesimal temperatures are considered, i.e. when states in the limit $T\to0^+$ are considered. 
We construct in this section a local order parameter $\bm{\Sigma}$ for the case at hand, that will be averaged over the full lattice, a non-zero value in the thermodynamical limit revealing an ordered, symmetry breaking phase. 
We recall here that several order parameter definitions are possible, and that specific order-parameters are required for different broken symmetry. 

When each local configuration can easily be associated with a ground-state, the order-parameter can take values in the ground state manifold, under some conditions on the broken symmetry, discussed below. 
This is the case for the local magnetization of ferromagnetic Ising or Heisenberg models, for example, where the order parameter is defined on each lattice site as the spin orientation, or for the alternated magnetization of N\'eel orders. 
In these cases, the order parameter takes values in $\mathcal S_2$ and can reveal a $\mathcal S_2$ symmetry breaking (at $T=0$, or in 3 dimensions for example). 

Complications arise when the definition of a ground-state involves several sites, with constraints on the spin orientations.
The antiferromagnetic triangular lattice is such an example: the sum of 3 spins of a triangle is zero at $T=0$, and the orientation of two non colinear spins are required to fully determine a ground state. This ground-state manifold is homeomorph to $SO(3)$\cite{Kawamura1985}.
For $T\neq0$, the constraint on the sum of spins is no more verified and there is no direct way to chose a ground state related to this configuration. 
We are here quite lucky, as a local configuration on a triangle of the kagome lattice can uniquely be propagated over the full lattice to form a state of the octahedral phase. 
A first possible order parameter is such a triplet of unit spins, forming an element of ${S_2}^3$. 
However, ${S_2}^3$ as order parameter space does not do the job to reveal a possible symmetry breaking. 
The Mermin-Wagner theorem states that continuous symmetries are unbroken at finite temperature. 
Here, there are the global spin rotations $SO(3)$. 
Thus $SO(3)$ forms classes of equivalence in ${S_2}^3$ such that at infinitesimal temperature, spin waves disorder the ground state and disperse the local order parameter over the full equivalence class, when measured over the full lattice. 
Each such class has a zero average in ${S_2}^3$, which rules out ${S_2}^3$ as order-parameter space to detect any finite temperature phase transition. 

Thus we are forced to use a $SO(3)$ invariant description of the ground-state manifold, as the quotient ${\mathcal S_2}^3 / SO(3)$, in order to appropriately account for the possible symmetry breakings.
Each point in ${\mathcal S_2}^3$ is defined by 6 parameters, while $SO(3)$ is a tridimensional manifold, from which we deduce that ${\mathcal S_2}^3/SO(3)$ has dimension 3 as well.
Points in this space, equivalence classes of states, must be described using $SO(3)$ invariants built from the initial variables $(\mathbf S_A, \mathbf S_B, \mathbf S_C)$ on a triangle $ABC$. 
An obvious choice is to use the dot product, which immediatly provides us with three invariants, that we group in a \textit{vector} $\bm\sigma(\mathbf S_A, \mathbf S_B, \mathbf S_C) = (\mathbf S_B\cdot \mathbf S_C, \mathbf S_C\cdot \mathbf S_A, \mathbf S_A\cdot \mathbf S_B)$. 
The image of $\bm\sigma$ is a subset of $\mathbf R^3$, whose shape is a slightly inflated tetrahedron. 
Its vertices correspond to colinear configurations, with three $\pm1$ vector components, and it can be shown that this shape indeed has the tetrahedral group $T_d$ as its symmetry group.
Note that we have lost the distinction between time-reversed spin configurations $\mathbf S_i\to - \mathbf S_i$. 
Points in the image of $\bm\sigma$ have one class of pre-images when the three spins are coplanar (as spin inversion is equivalent to a rotation of $\pi$ in this case), two when they are not. 
Thus, $\bm\sigma$ is unable to describe the breaking of the $\mathbb{Z}_2$ inversion subgroup of the $O(3)$ global spin transformations. 

\begin{figure}
    \begin{center}
        \includegraphics[width=0.44\textwidth]{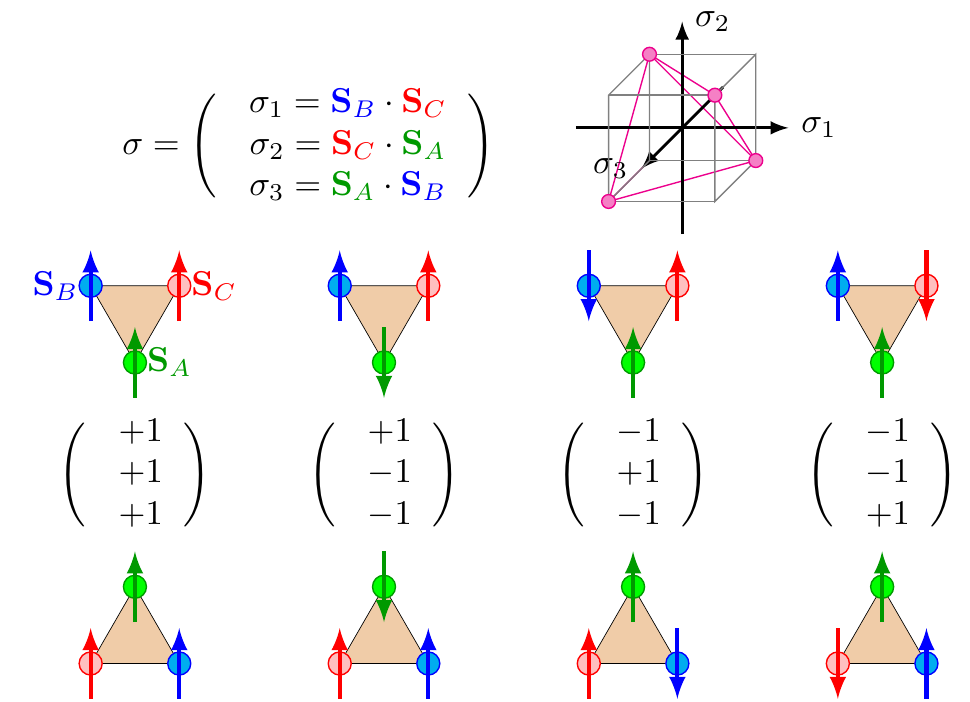}
        \caption{Four possible spin configurations
            for $T\to0^+$ on a reference triangle of the kagom\'e lattice 
            (up to a global spin rotation).
            They are labeled by a triplet of spin dot products $\bm\sigma$ and stay on the vertices of a tetrahedron. }
        \label{fig:four_potts_orders} 
    \end{center}
\end{figure}

Returning to ObD, the alignement of all spins can now be easily identified using $\bm\sigma$. 
The tendency to colinearity of neighboring spins can be visualized as free energy barriers
effectively pushing the ground-state configurations towards the vertices of the inflated tetrahedron, points of high symmetry, describing perfect (anti)-alignment in spin triplets. 
By considering vertices only, one can quickly observe that each vertex is invariant under the permutation of the three others, $\mathcal S_3$, while the whole symmetry group is isomorphic to the permutation group of four points $\mathcal S_4$. 
Consequently, our points may be described as the quotient space $\mathcal S_4/\mathcal S_3 \simeq K_4$, a genuine group since $\mathcal S_3$ is normal in that case. 
This group provides the set of transformations that allows us to navigate between the different colinear ground states, by flipping pairs of spins (or not flipping any for the neutral element), and is thus the actual symmetry broken by this phase transition - they simply represent the action of translations of the lattice on a ground state.
As a time-reversal spin transformation ($\mathbf S_i\to - \mathbf S_i$) let the elements of this group invariant, the impossibility to distinguish states breaking this symmetry, evocated above, does not evince $\bm{\sigma}$ as an appropriate order parameter. 

Up to now, we have considered a single reference triangle $ABC$. 
Depending on the choice of the labels $A$, $B$ and $C$ of the triangle vertices (4 possibilities), $\bm\sigma$ undergoes a transformation. 
To fix the definition of $\bm\sigma$, its $i$th component $\sigma_i$ is defined as the dot product of spins on a link directed along the vector $\mathbf e_i$ of Fig.~\ref{fig:Potts_order}. 
This unambiguously defines $\bm\sigma$ on all the pointing-down as well as pointing up triangles (see Fig.~\ref{fig:four_potts_orders}). 

\begin{figure}
    \begin{center}
        \includegraphics[width=0.45\textwidth]{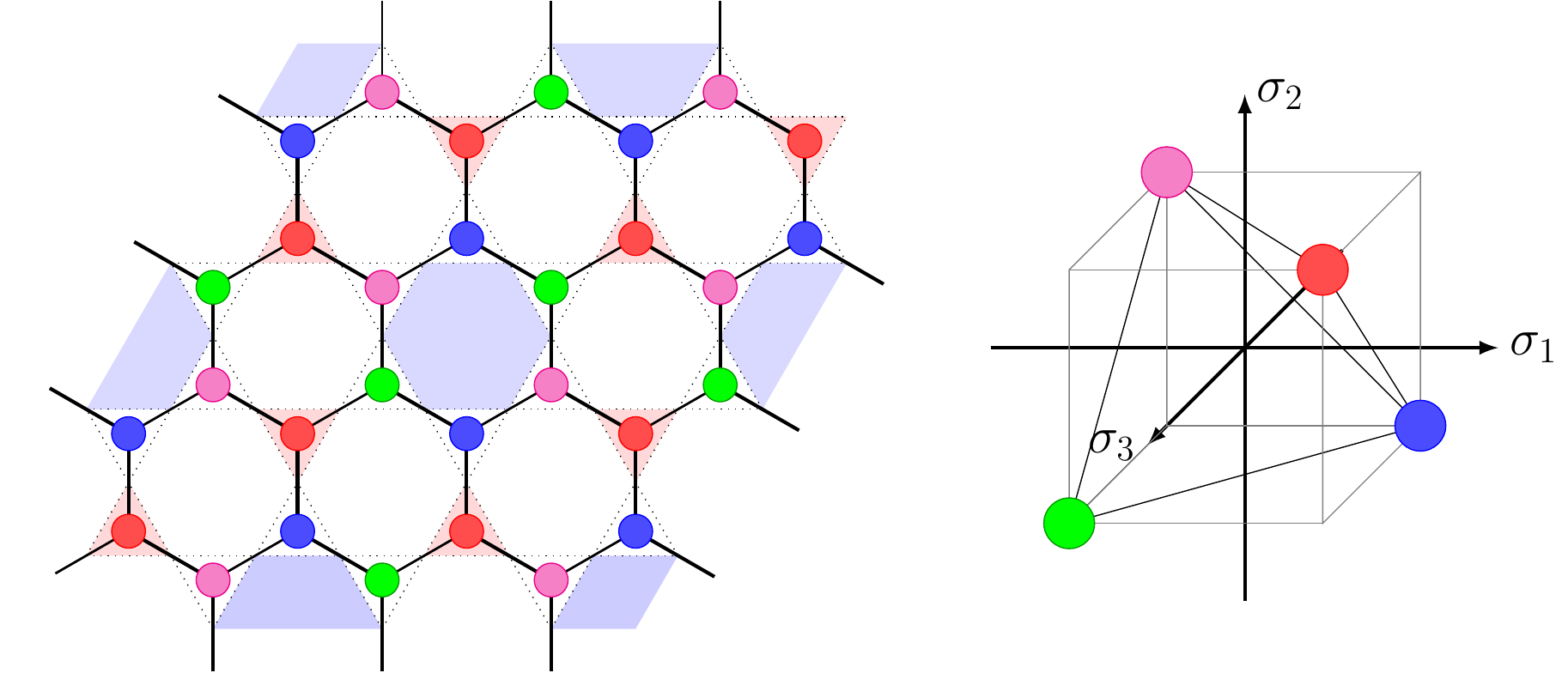}
        \caption{Lattice of effective spins, i.e. values of the tridimensional order parameter locally defined on each triangle. Shared vertices between triangles are represented as edges linking the corresponding sites. 
        Each color corresponds to a value a given value of $\bm\sigma$, as shown on the right panel.}
        \label{fig:hexa_lattice} 
    \end{center}
\end{figure}

The four possible triplets for colinear configurations are represented on Fig.~\ref{fig:four_potts_orders}.
The centers of up and down triangles on the kagom\'e lattice form a honeycomb lattice, and $\bm\sigma$ is an effective (non unit) spin on these sites, oriented alternatingly as indicated on Fig.~\ref{fig:hexa_lattice} in a colinear ground state configuration. 
Note that once $\bm\sigma$ is chosen on one of the kagom\'e triangle in a colinear ground state configuration (or equivalently on one of the honeycomb lattice sites), $\bm\sigma$ on any other triangle can be deduced from elementary operations belonging to the Klein group $K_4$: 
an $\mathbf e_i$ translation of the spins rotates $\bm\sigma$ by $\pi$ around the $\sigma_i$ axis. 
The tetrahedra of  $\bm\sigma$ orientations falls in one of four possible orientations, corresponding to a $q=4$ Potts variable\cite{Wu1982}. 

By analogy with the alternate order parameter used for antiferromagnetic long-range order, we define an alternate order parameter $\bm\Sigma$, homogeneous over the full lattice.  
The evolution of its average over the full lattice as a function of the temperature and of the system size will now be studied below using Monte Carlo simulations. 
Note that in a colinear ground state, $\bm\Sigma$ is homogeneous, and only four ground states are possible. 
In this aspect, the effective model for the $\bm\Sigma$ variables ressembles more to the ferromagnetic $q=4$ Potts model than to the antiferromagnetic one, whose degeneracy on the honeycomb lattice would be extensive.

\section{Monte Carlo simulations at finite temperature}
\label{sec:MC}

\subsection{The method}
\label{sec:MC_method}
To investigate the phase diagram of the $J_1-J_3$ model, we perform Monte Carlo simulations by implementing a parallel-tempering method\cite{Bittner2011}. 
In the case of first-order phase transitions, this method enables to overcome the associated free-energy barriers by considering $N_p$ replicas of the system at different temperature $T_i$, with $i=1, \dots, N_p$. 
Each replica constitutes a separate, parallel, simulation box whose state evolves independently via local spin updates, but can also periodically be swapped with that of its immediate neighbors. 
Hence, higher temperature simulation boxes allow lower temperature ones to sample their phase space much more efficiently.
The temperature interval is chosen in order to cover the region where a putative phase transition is expected, and the difference of inverse temperature between two adjacent replicas $\Delta\beta$ is kept constant 
(we also also tried a geometric progression for the inverse temperatures in the range, without noticing significant changes for the convergence of the method). 

In order to satisfy a detailed balance for this process, the probability $P_{PT}$ of accepting an exchange of configurations between boxes $i$ and $i+1$ is chosen with a Metropolis rule
\begin{equation}
P_{PT}(i\leftrightarrow i+1)={\rm Min}(1,\exp(\Delta\beta \Delta E)),
\end{equation}
with $\Delta \beta=\beta_i-\beta_{i+1}$ and $\Delta E=E_i-E_{i+1}$. 
The double arrow means that the probability $P_{PT}$ is symmetric to the reverse exchange.

The mean acceptance probability $P_A(i\leftrightarrow i+1)$ between boxes $i$ and $i+1$ is the average of $P_{PT}(i\leftrightarrow i+1)$ over thermalized configurations, and writes:
\begin{eqnarray}
\label{eq:prob_accept}
P_A(i\leftrightarrow i+1)&=&\int dE_i\,dE_{i+1}
\\
&& P_{\beta_i}(E_i)
P_{\beta_{i+1}}(E_{i+1})P_{PT}(i\leftrightarrow i+1),
\nonumber
\end{eqnarray}
where $P_{\beta_i}(E_i)$ denotes the equilibrium probability of the box $i$ to have an energy $E_i$. 
Eq.~\eqref{eq:prob_accept} is merely a weighted sum over all possible energetic configurations for two given neighboring boxes.
In order to optimally schedule the temperatures, we check that the acceptance probability 
of swaps between neighboring replicas is near $0.5$\cite{Bittner2011}. 

We choose an even number of replicas $N_p$ and at constant time intervals, two kinds of exchanges between neighboring boxes are proposed:
either exchanges between all pairs $(2k-1,2k)$ where $k=1, ..., N_p/2$ or exchanges between all pairs $(2k,2k+1)$ where $k=1, ..., N_p/2-1$, which preserves the ergodicity of the process.
Otherwise, we perform local updates of spins for each simulation box according to a Metropolis rule.  

In simulations on a lattice of size $L$, we store the histograms of the energy and of the order parameter modulus $|\sum_{\triangledown, \vartriangle} \bm\Sigma|$ for each temperature, giving directly access to the mean energy $\langle E\rangle(\beta,L)$, 
and the mean Potts magnetization $\langle \Sigma \rangle(\beta,L)$. 
The specific heat $C_V$, the susceptibility of the order parameter $\chi_\Sigma$, and the associated Binder parameter $B_\Sigma$ are given per lattice site as:
\begin{subeqnarray}
C_V(\beta,L)
&=& \frac{\beta^2}{N}\left(\langle E^2\rangle - \langle E\rangle^2\right)\\
\chi_\Sigma(T,L)
&=&  N (\langle \Sigma^2\rangle - \langle \Sigma\rangle^2)
\\
B_\Sigma(\beta,L)&=& 1- \frac{\langle \Sigma^4\rangle}{3\langle \Sigma^2\rangle}.
\end{subeqnarray}

Moreover, by using the reweighing method\cite{Ferrenberg1988},and the histograms obtained in simulations,
one builds  for each box $i$ all estimated above quantities within
a temperature interval  $[(\beta_i+\beta_{i-1})/2,(\beta_i+\beta_{i+1})/2]$.
Collecting all curves, one can build a global graph from $T_{\rm min}$ to 
$T_{\rm max}$. The convergence for all temperatures of the parallel tempering method is confirmed when the curve is continuous at each boundary between two temperature intervals.

In order to perform a finite scaling analysis, we simulated different system sizes of the kagom\'e lattice with periodic boundary conditions. 
$L$ is the linear size of the lattice, and the number of sites if $N=3L^2$. 
By using simulation data, we determine  the maxima $C_V^{\rm max}(L)$ and $\chi_\Sigma^{\rm max}(L)$ of these quantities, 
occurring at temperatures $T_c^{C_V}(L)$ and $T_c^{\chi_\Sigma}(L)$.
For a continuous phase transition, the finite size scaling at the lowest order of these quantites is given by:
\begin{subeqnarray}
C_V^{\rm max}(L) &\simeq& a L^{\alpha/\nu}+b,
\\
\chi_\Sigma^{\rm max}(L)&\simeq& c L^{\gamma/\nu}+d,
\\
T_c^{C_V,\chi}(L)&\simeq& eL^{-1/\nu} + T_c(\infty), 
\end{subeqnarray} 
where $\alpha$, $\nu$ and $\gamma$ are critical exponents whose values for the ferromagnetic $q=4$ Potts model are recalled in App.~\ref{App:exponents_q4} and $T_c(\infty)$ is the critical temperature of the phase transition. 
For a first-order phase transition, the finite size scaling is given by:
\begin{subeqnarray}
C_V^{\rm max}(L) &\simeq & a L^{2}+b,
\\
\chi_\Sigma^{\rm max}(L)&\simeq& c L^{2}+d,
\\
T_c^{C_V,\chi}(L)&\simeq& eL^{-2} + T_c(\infty).
\end{subeqnarray} 
\begin{figure}
    \begin{center}
        \includegraphics[width=0.45\textwidth]{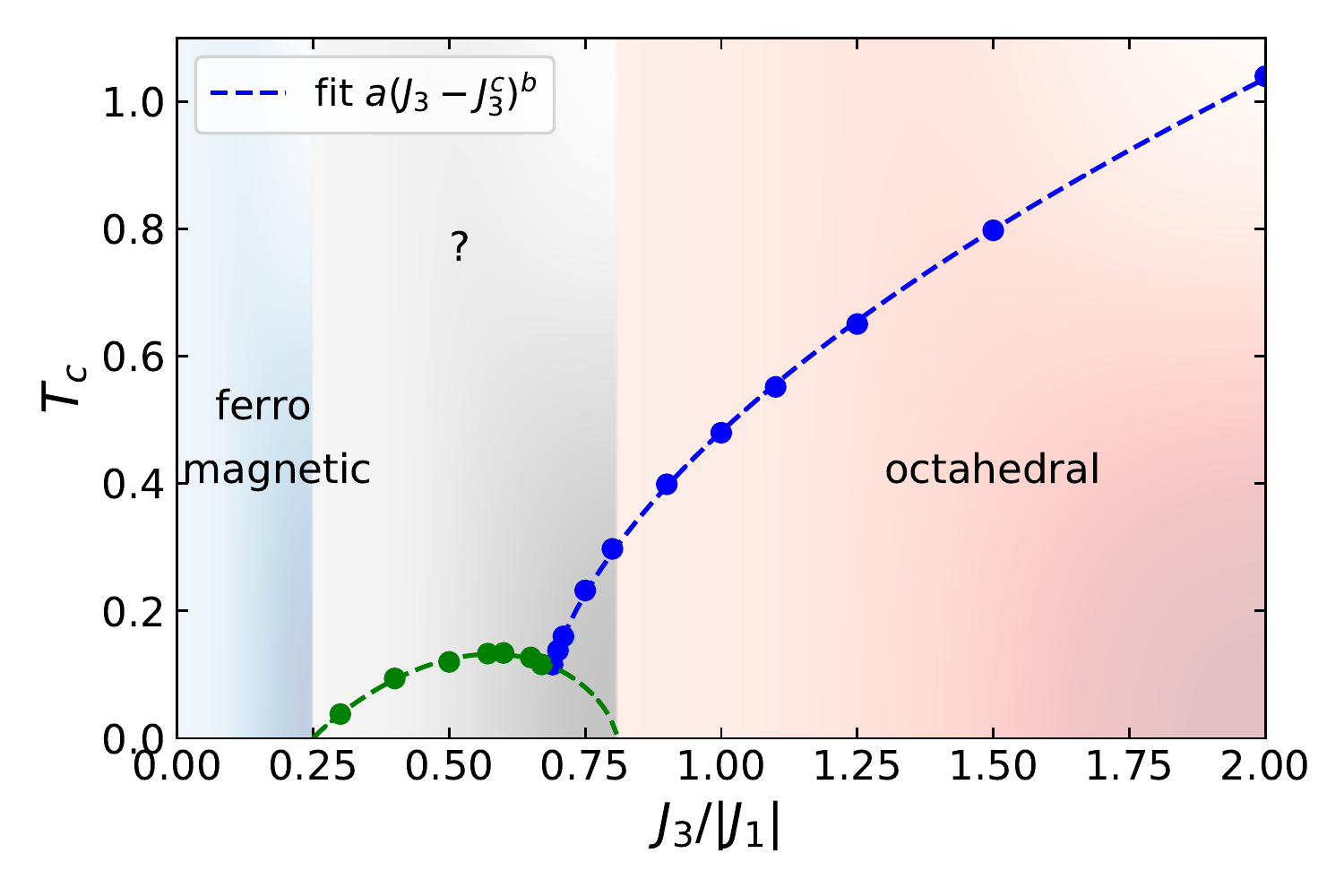}
        \caption{Phase diagram of the $J_1-J_3$ Heisenberg model on the kagom\'e lattice. 
        A phase transition with both $C_V$ and $\chi_\Sigma$ divergency (blue points) is evidenced by Monte Carlo classical simulations, restoring the $K_4$ symmetry. 
        The dashed blue line is a 3-parameter fit with a power law: $a=0.88$,  $b=0.56$,  $J_3^c=0.66$. 
        Green points are phase transitions with no $\chi_\Sigma$ divergency. 
        The green dashed line is a guide to the eyes. }
        \label{fig:phase} 
    \end{center}
\end{figure}

\subsection{Results for ferromagnetic $J_1$}

The linear size of the lattice $L$ goes from $12$ to $104$.
The interaction between nearest neighbors $J_1$ is set to $-1$ and $J_3$ is varied from $0.2$ to $2$. 
By considering the $T=0$ phase diagram (top of Fig.~\ref{fig:camembert}), this corresponds to a vertical line in the upper left quarter, which intersects three ground state sectors: ferromagnetic, unconventional and octahedral. 
One leaves the ferromagnetic phase when $J_3=\frac 14$ and enter the degenerate octahedral phase for $J_3 = \frac{1+\sqrt{5}}{4} \simeq 0.809$, where one expects a finite temperature phase transition due to emergence of the discrete $K_4$ order parameter.
Note that the table \ref{tab:phi} gives an one-to-one mapping between the coupling ratio $J_3/|J_1|$ and the parameter $\phi$ introduced in the preceding section.

\begin{table}
    \begin{tabular}{|c|c|c|c|c|c|c|c|}
        \hhline{--|~|--|~|--}
        $J_3/|J_1|$&$\phi/\pi$&&$J_3/|J_1|$&$\phi/\pi$&&$J_3/|J_1|$&$\phi/\pi$\\
        \hhline{|=|=|~|=|=|~|=|=|}
        0  & 1 &&	0.68 &0.810	&& 1 & 3/4
        \\
        \hhline{--|~|--|~|--}
        0.25& 0.922 && 0.75 & 0.795  && 2 & 0.648
        \\
        \hhline{--|~|--|~|--}
        0.5 &0.852 &&	0.809 &0.783 && $\infty$ & 1/2 
        \\
        \hhline{--|~|--|~|--}
    \end{tabular}\label{tab:phi}
    \caption{$J_3/|J_1|$ versus $\phi$ for $J_1=-1$. 
        $\phi_F \simeq 0.922\pi$ and $\phi_O \simeq 0.783\pi$ are the boundaries of the unconventional phase, whose exact value is given in Fig.~\ref{fig:camembert}. }
\end{table}

$C_V$ and/or $\chi_\Sigma$ shows an maximum increasing with $L$ for some $J_3$ values, revealing a phase transition. 
The resulting finite temperature phase diagram is displayed in Fig.~\ref{fig:phase}. 
Blue points indicate both a $C_V$ and $\chi_\Sigma$ divergency, whereas green points indicate that only $C_V$ diverges. 

We now discuss in more detail our results by considering the three different regions (ferromagnetic, unconventional and octahedral ground states). 

\subsubsection{Ferromagnetic region: no transition}
For $J_3=0.2$ (let us recall that $J_1$ is set to $-1$ in simulations), no phase transition was observed, at any temperature. 
There is no evolution of the specific heat with the system size. 
Hence our results are in line with the predictions of the Mermin-wagner theorem for this phase, as expected.

\begin{figure*}
    \begin{center}
        \includegraphics[width=\textwidth]{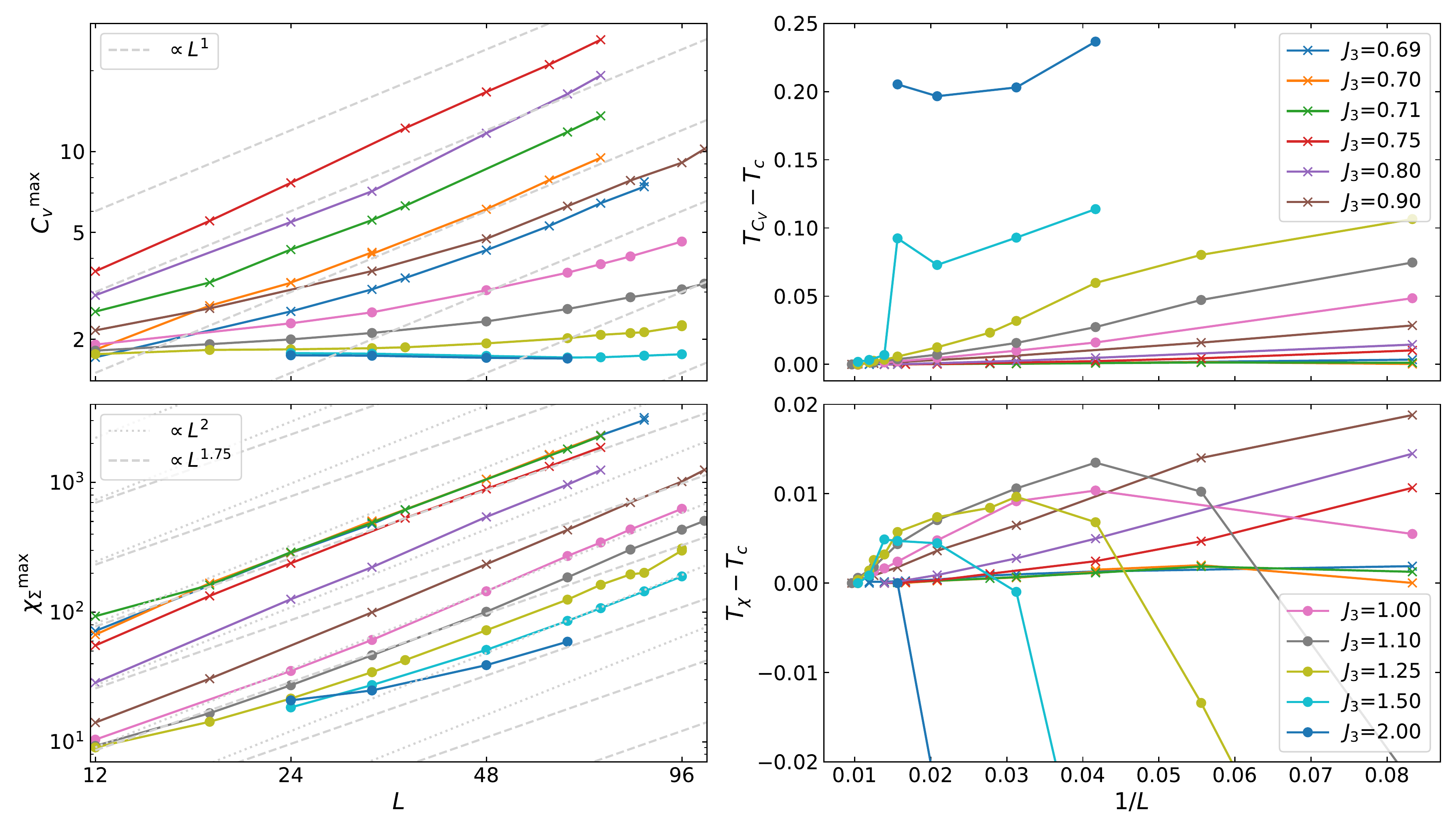}
        \caption{Maximum of $C_V$ and of $\chi$ versus the lattice size $L$ for $J_1=-1$ and various $J_3$, and temperature of their maxima. $T_c$ has been extracted from $\chi_\Sigma$. }
        \label{fig:maxCvChi} 
    \end{center}
\end{figure*}

\subsubsection{Non $K_4$ phase transitions in the unconventional phase}
When $0.25<J_3<0.809$, the ground state is not easily determined and seems to be very dependent of $J_3$, as explained in Sec.~\ref{sec:model_and_GS} (for example, with a succession of various types of wave vectors). 
The following values of $J_3$ have been explored: 0.3, 0.4, 0.5, 0.6, 0.65, 0.67, 0.69, 0.7, 0.71, 0.75, 0.8, all showing a unique divergency of $C_V(\beta, L)$ with $L$. 

For $0.25 \leq J_3\leq0.67$, the $K_4$ Potts parameter $\Sigma$ remains close to zero at all temperatures. 
However, the specific specific heat displays a peak at low temperature, whose size increases with $L$. 
The approximative limit of $T_{C_V}$ when $L$ increases seems to be a continuous function of $L$ and is indicated as green points on Fig.~\ref{fig:phase}: it increases from zero for $J_3=1/4$ up to $T_c = 0.134(1)$ for $J_3\simeq 0.60(3)$, and slightly decrease down to $0.116(1)$ up to $J_3=0.67(2)$. 
Due to the nature of the ground state, it is possible that transitions associated with various broken symmetries occur in this range of parameters. 
It is for example probable that the three-fold spatial rotation is broken at low $T$ for $J_3\simeq 0.67$ as the order of Fig.~\ref{fig:config_unconventional} particularizes one of the three sublattices. 
We did not try to identify the order parameter associated with these phase transitions as the focus of this study is the octahedral phase. 

For  $0.6\leq J_3\leq 0.67$, the mean energy per site at low $T$ depends on the system size even quite far from the critical temperature. 
Moreover the temperature of $C_V^{\rm max}$ varies non monotonously with the system size. 
These features are the signature of a phase transition twarted by the incommensurability of the lattice size with the periodicity of the order, inducing frustration.
The phenomenon weakens when $L$ increases, and could be handled using twisted boundary conditions. 

Lastly, the energy distribution is unimodal for $J_3<0.6$, but becomes bimodal for system sizes of $L\geq 32$ (24) and $J_3=0.6$ (0.65), which is in favor of a first-order phase transition. 
For $J_3=0.67$, the energy distribution consists in two well separated peaks near $T_c$, even at low $L$ and the phase transition is clearly first order.

For $0.69\leq J_3 \leq 0.809$, $\Sigma$ has large values in the low $T$ phase and its susceptibility shows a peak which increases with $L$. 
For this reason, this transition will be discussed in the next paragraph, on the $K_4$ transition. 
Such $K_4$ transition is surprising here as the $T\to0$ state is not supposed to break the $K_4$ symmetry: $\Sigma$ should be zero in the non-octahedral ground state. 
Another phase transition thus seems unavoidable at lower $T$, restoring $K_4$.
In this hypothesis, the green dashed line of Fig.~\ref{fig:phase} was extended up to 0.809, implying a reentrance of the $K_4$-breaking phase in the unconventional phase.
The low-$T$ phase transition would be first-order, as it relates phases with different broken symmetries. 
However, we did not succeed to evidence such a low-$T$ phase transition, probably because of metastable states breaking $K_4$, in which the simulations remains stucked despite the parallel tempering.

\subsubsection{$K_4$ phase transition, in the unconventional and octahedral regions}

For $J_3\geq0.69$, a transition occurs with both a $C_V^{\rm max}$ and a $\chi_\Sigma^{\rm max}$ divergency with $L$, occuring at temperatures converging towards the same value $T_c(J_3)$.
$C_V^{\rm max}$ and of $\chi_\Sigma^{\rm max}$ have been collected for various of $L$ and $J_3$ on Fig.~\ref{fig:maxCvChi}, together with their temperatures.
Finally, the Binder cumulant associated with $\Sigma$ displays the behavior associated with a phase transition: it tends to 2/3 below $T_c$ when $L$ increases and its curves for different $L$ cross at the same temperature. 
This transition separates a low-$T$ phase with large $\Sigma$ from a nearly zero $\Sigma$ high-$T$ one. 
It corresponds to the restoration of the $K_4$ Potts symmetry, at a temperature $T_c(J_3)$ that increases with $J_3$ (Fig.~\ref{fig:phase}). 
$T_c(J_3)$ is well fitted by $a (J_3 - J_3^c)^b$, with the three adjustable parameters $a=0.88$,  $b=0.56$ and  $J_3^c=0.66$.
We have here the proof that the order by disorder favors colinear states among the ground state manifold at low temperature. 

\begin{figure}
	\begin{center}
		\includegraphics[width=0.45\textwidth]{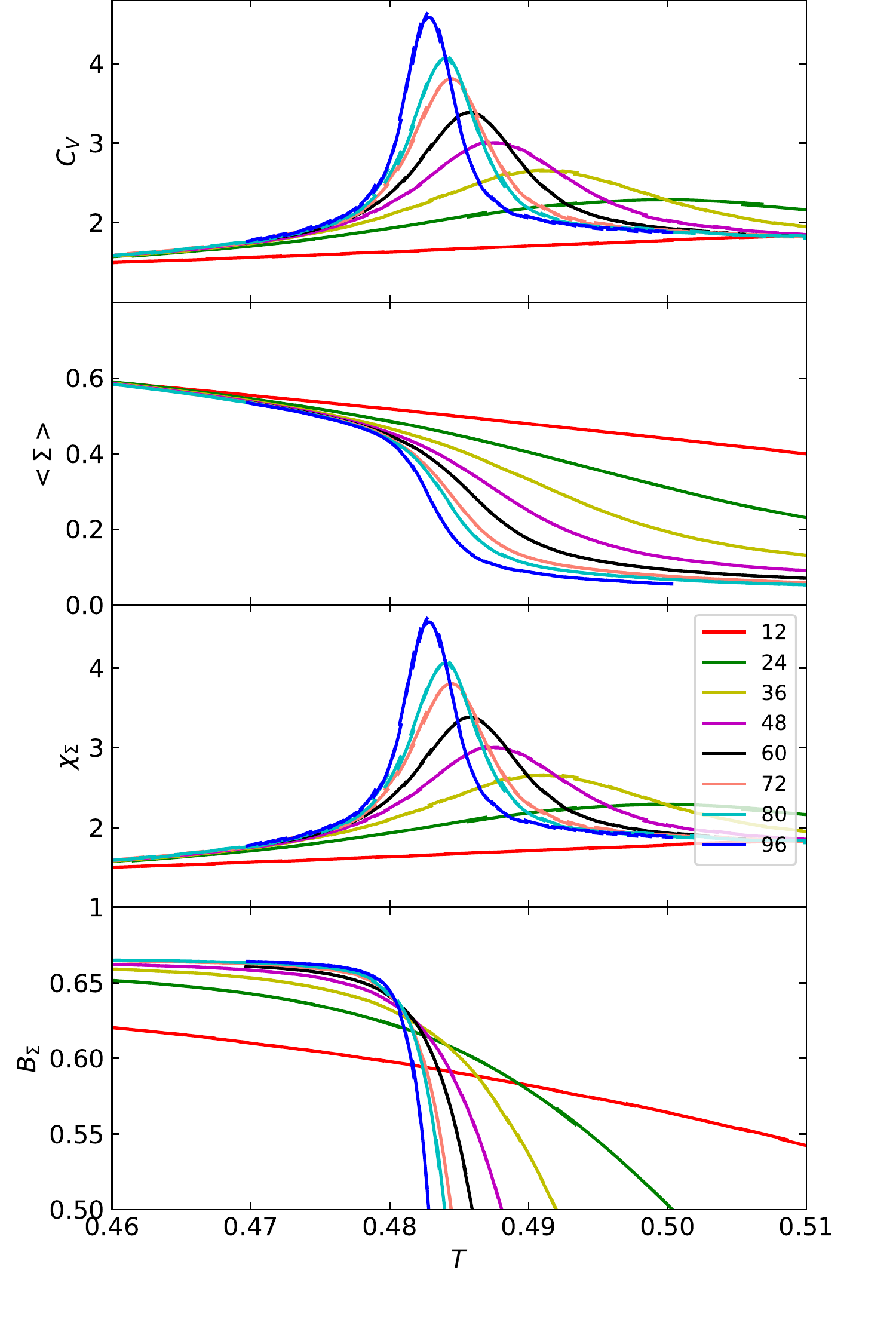}
		\caption{Specific heat $C_V$, $K_4$ order parameter $\Sigma$, susceptibility $\chi_\Sigma$ and Binder parameter $B_\Sigma$
versus $T$ for different system sizes $L$ for $J_1=-1$ and $J_3=1$. }
		\label{fig:J31} 
	\end{center}
\end{figure}

Fig.~\ref{fig:J31} shows several quantities (Specific heat $C_V$, Potts magnetisation $\Sigma$, susceptibility $\chi$ and Binder parameter $B$) as a function of $T$ for different system sizes, for $J_3=1$, 
as an illustration of a finite size scaling. 

At low $J_3 \lesssim 1$, the energy distribution is weakly bimodal near $T_c$, which means that the two peaks are not well separated at low $L$. 
Both $C_V$ and $\chi_\Sigma$ show a nice divergency, at a temperature that extremely rapidly converges (Fig.~\ref{fig:maxCvChi}), making the determination of the exponents related to it unpossible due to precision issue. 
The exponents of the growth of $\chi_\Sigma$ is very near 2, which supports the hypothesis of a first order transition, but the one for $C_V$ remains near $1$, against 2 expected.
It may be as a consequence of the unclear separation of the two peaks in the energy distribution, revealing a finite, but very large correlation length at the critical temperature, that would require simulations with larger lattice size. 
Another explanation would be that the transition becomes second ordered. 
Then, if it is in the universality class of the $q=4$ Potts model, the exponents should be $\alpha/\nu=1$ and $\gamma/\nu=7/4$. 
These values are possible, but cannot be confirmed in view of our calculations. 

The energy distribution at $T_c$ becomes unimodal for $J_3\gtrsim 1$ up to the explored lattice sizes. 
Together with this change, the maximum of the specific heat needs much larger lattice sizes to convincingly increase with $L$ (Fig.~\ref{fig:maxCvChi}). 
This is more and more pronounced when $J_3$ increases: for $J_3\gtrsim 1.25$, we even see the appearance at large size of a secondary peak in $C_V$, that develops itself on the side of the main broad peak. 
For $J_3=1.5$, it only catches up the broad-peak maximum value at $L\simeq 64$, as can be seen on Fig.~\ref{fig:maxCvChi}, where it translates in a dropout of $T_{C_V}$ with $L$. 
It becomes tedious to extract critical exponents for $C_V$ because the prefactor of the scaling behavior is very small.
The signature of the transition is still present in the scaling behavior of the order parameter: $\chi_\Sigma$ displays clear sign of divergency, even at small lattice sizes, with an exponent that remains near 2. 

To conclude, we observe a phase transition for $J_3>0.69$ associated with $\Sigma$, that is weakly first order for small $J_3$. 
With increasing $J_3$, the first order transition still weakens, up to a point where it could be a second order transition. 
However, the critical exponents are difficult to determine due to the large sizes required to observe the leading order behavior of the maximum of $C_V$, but could correspond to those of the $q=4$ Potts model. 
In the case of the antiferromagnetic $J_1-J_2$ square lattice, where order by disorder tends to align spins for $J_2>J_1/2$, the same difficulty was observed\cite{Weber2003} when the sublattices become less coupled (when $J_2$ increases for the square lattice, $J_3$ for the kagom\'e). 

\subsection{Results for antiferromagnetic $J_1$}

\begin{figure}
    \begin{center}
        \includegraphics[width=0.47\textwidth]{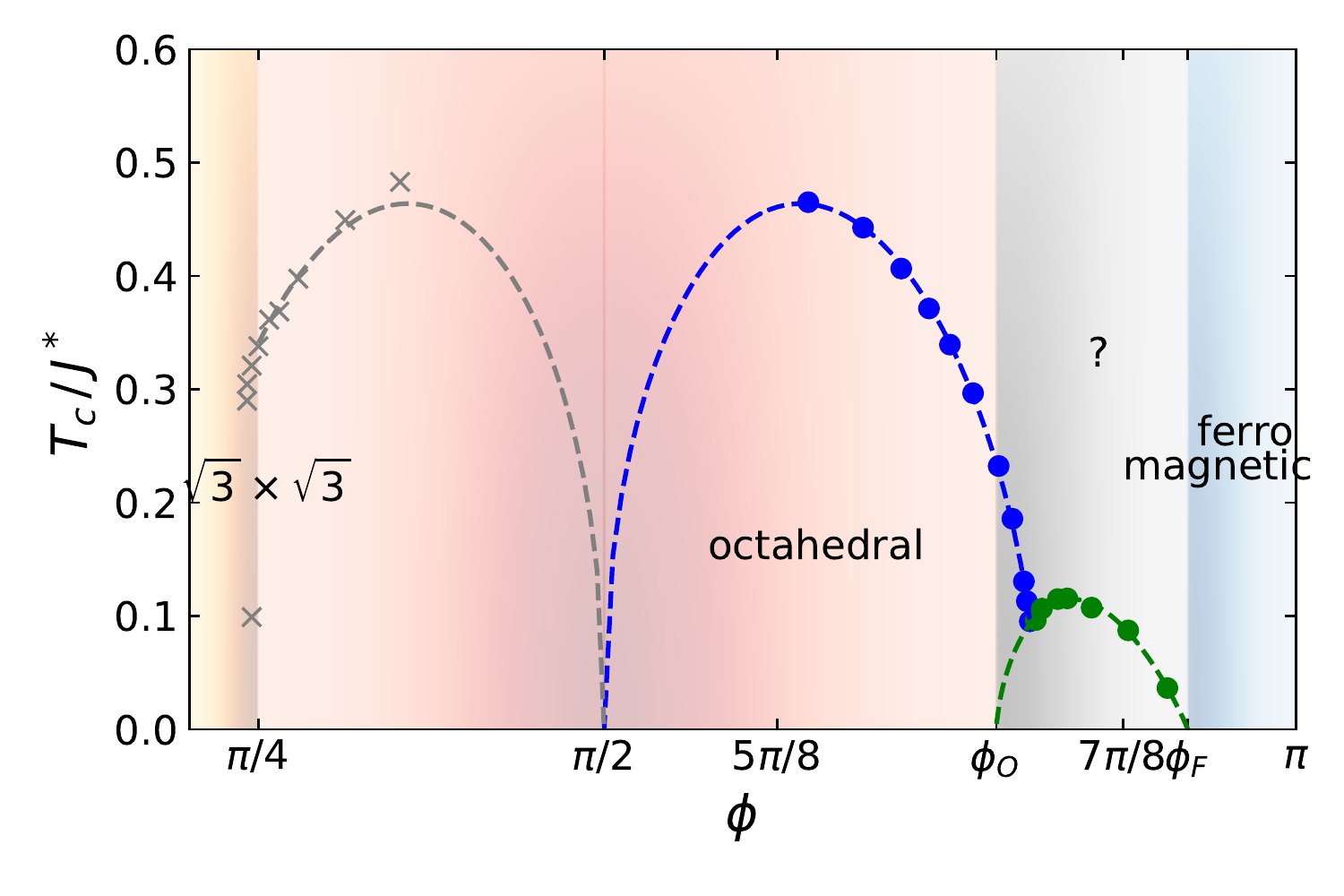}
        \caption{Phase diagram of the $J_1-J_3$ Heisenberg model on the kagom\'e lattice as a function of $\phi$. 
        Phase diagram obtained by Monte-Carlo simulation of the $J_1-J_3$ model,in the $T-\phi$ plane. 
        The blue and green points are those of Fig.~\ref{fig:phase}, for $J_1<0$, with the same fits. 
        Grey points are obtained for $J_1>0$, and the grey dashed line is obtained by symmetry of the blue one with respect to $\phi=\pi/2$. 
        }
        \label{fig:phase_phi} 
    \end{center}
\end{figure}

In order to explore the full octahedral phase of the phase diagram, we have also investigated the model with an antiferromagnetic interaction between the first nearest spins ($J_1=1$).
However, this situation is not supposed to describe the Ba-Vesignieite compound. 
Simulation are performed for various positive values of $J_3$ and the transition temperatures are displayed on Fig.~\ref{fig:phase_phi}, which translates Fig.~\ref{fig:phase} in terms of $\phi$ and extends it to positive $J_1$ values. 
A astonishing similarity with the ferromagnetic $J_1$ is found: the transition temperature does not depend on the sign of $J_1$ for $J_3>1$, as emphasized on Fig.~\ref{fig:phase_phi}. 
It suggests that the critical temperature is only a function of $\sin\phi$ in a large neighborhood of $\phi=\pi/2$. 
Note that as soon as the leading term of $T_c(J_3)$ is weaker than $J_3/|J_1|$, $\lim_{\phi\to \pi/2} T_c(\phi) = 0$, which seems coherent as in this limit, the three sublattices are completely independent and no order is expected, at any temperature. 

For $J_3<1$, the ground state is in the $\sqrt3\times \sqrt3$ phase and at low $T$, $\Sigma$ is effectively very low. 
However, for $0.95<J_3<1$, it sharly increases above a first critical temperature, and goes down again at a second one. 
This shows the existence of a reentrance of the $K_4$ symmetry broken phase in the $\sqrt3\times \sqrt3$ phase. 
This behavior is here more easily detected than in the unconventional phase, where it was only conjectured. 
This is probably due to the nature of the $\sqrt3\times \sqrt3$ low $T$ phase, that here does not break any symmetry and must cause less thermalization issue.

\section{Quantum and thermal fluctuations: linear spin wave approximation}
\label{sec:colinear_selection}

An analytical approach to understand the emergence of 
a discrete order parameter, leading to a phase transition, consists in 
departing from one of the classical ground states, which are all 
equally favored at strictly zero temperature in the classical model, 
and to perturb it by adding infinitesimal thermal or quantum fluctuations 
(perturbing the classical state either by an infinitesimal $T$ or $1/S$). 
We thus expect to lift the degeneracy between them. 
Thermal and quantum perturbations can be apprehended through the same formalism called the linear spin wave approximation. 
It will be developed in the two next subsections. 
But let us first develop the part which is common to both perturbations and define a set of eigenenergies $\omega_{\mathbf q, l}$ which will be exploited differently in each case. 

First, a reference ground state is chosen, whose spin orientation on site $i$ is $\mathbf S_i^0$. 
We then chose a rotation $R_i$ such that $R_i \mathbf S^0_i = \mathbf e_z$ and label by $\mathbf S'_i$ the spin in the newly defined basis: $\mathbf S'_i=R_i \mathbf S_i$, whatever its orientation.
$\mathbf S'_i$ is either a real vector in the classical case, or an operator vector in the quantum case. 
In both cases, its norm is constrained by the spin length $S$. 
Using ${S'_i}^{\pm} = {S'_i}_x \pm i {S'_i}_y$, a vector $\mathbf U_i$ is defined as:
\begin{equation}
\mathbf U_i 
=
\begin{pmatrix}
{S'_i}^+
\\
{S'_i}^-
\\
{S'_i}^z
\end{pmatrix}
=
V \mathbf S'_i
\qquad 
V=
\begin{pmatrix}
1&i&0\\
1&-i&0\\
0&0&1
\end{pmatrix}
\end{equation}
The Hamiltonian written in terms of $\mathbf U_i$ is:
\begin{equation}
\label{eq:Ham_SW}
H=
\frac 12\sum_{i,j} \mathbf U_i \cdot \underbrace{(V R_i J_{i,j} R_j^{-1} V^{-1})}_{M_{i,j}}\mathbf U_j
\end{equation}
We now expand the Hamiltonian with respect to a small parameter related to the distance of the actual state with the reference ground state: $S - {S'}^z_i$. 
We need here to focus successively on the low-$T$ classical case and on the zero-$T$ quantum case, to finally get the same eigenmodes in both situations. 

To describe the quantum ground state, a Holstein-Primakoff transformation of the $\mathbf S_i'$ spins is performed.
It defines $a_i^\dag$ and $a_i$ bosonic creation and annihilation operators on each site $i$.
They are subject to a constraint on their number $n_i = a_i^\dag a_i \leq 2S$, to respect the spin length. 
$n_i$ is supposed to be $\mathcal O (1)$ in $S$: 
\begin{equation}
\label{eq:Uquant}
    \mathbf U_i
    =
    \begin{pmatrix}
        \sqrt{2S-a_i^\dagger a_i}\, a_i
        \\
        a_i^\dagger \sqrt{2S-a_i^\dagger a_i}
        \\
        S - a_i^\dagger a_i
    \end{pmatrix}
    =
    \begin{pmatrix}
        \sqrt{2S} a_i + \mathcal O\left(S^{-1/2}\right)
        \\
        \sqrt{2S} a_i^\dag + \mathcal O\left(S^{-1/2}\right)
        \\
        S-a_i^\dag a_i
    \end{pmatrix}
\end{equation}
The Hamiltonian now describes interacting bosons on the lattice.

On the classical side, by chosing as small complex parameter $z_i = \frac{{S'}_i^+}{\sqrt{2S}}$ and supposing it in $\mathcal O(1)$ (which is unjustified, as explained below), we get:
\begin{equation}
\label{eq:Uclass}
\mathbf U_i=
\begin{pmatrix}
\sqrt{2S}z_i\\
\sqrt{2S}z_i^*\\
\sqrt{S^2 - 2S|z_i|^2}
\end{pmatrix}
=
\begin{pmatrix}
\sqrt{2S}z_i\\
\sqrt{2S}z_i^*\\
S - |z_i|^2 + \mathcal O\left(S^{-1}\right)
\end{pmatrix}.
\end{equation}

\begin{figure*}
    \begin{center}
        \includegraphics[height=.22\textwidth]{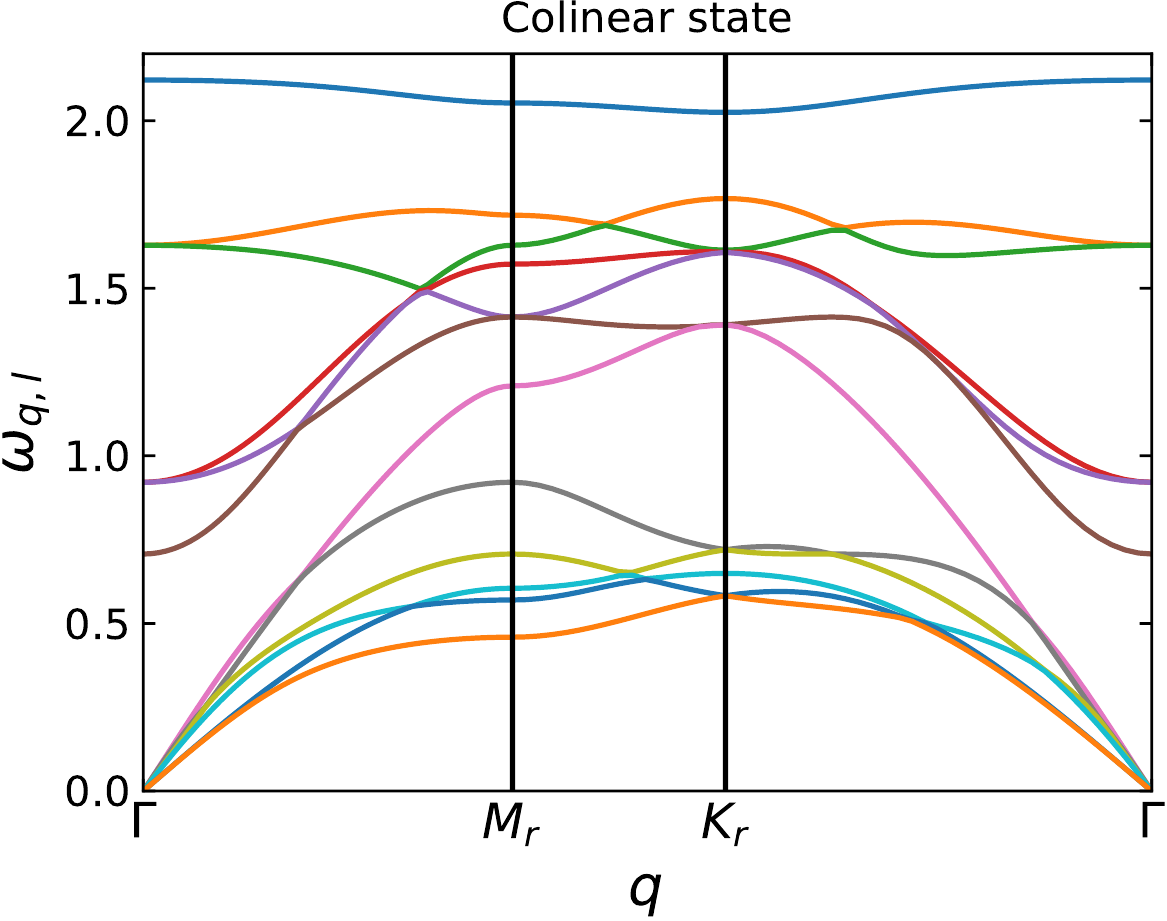}\quad
        \includegraphics[height=.22\textwidth, trim= 38 0 0 0, clip]{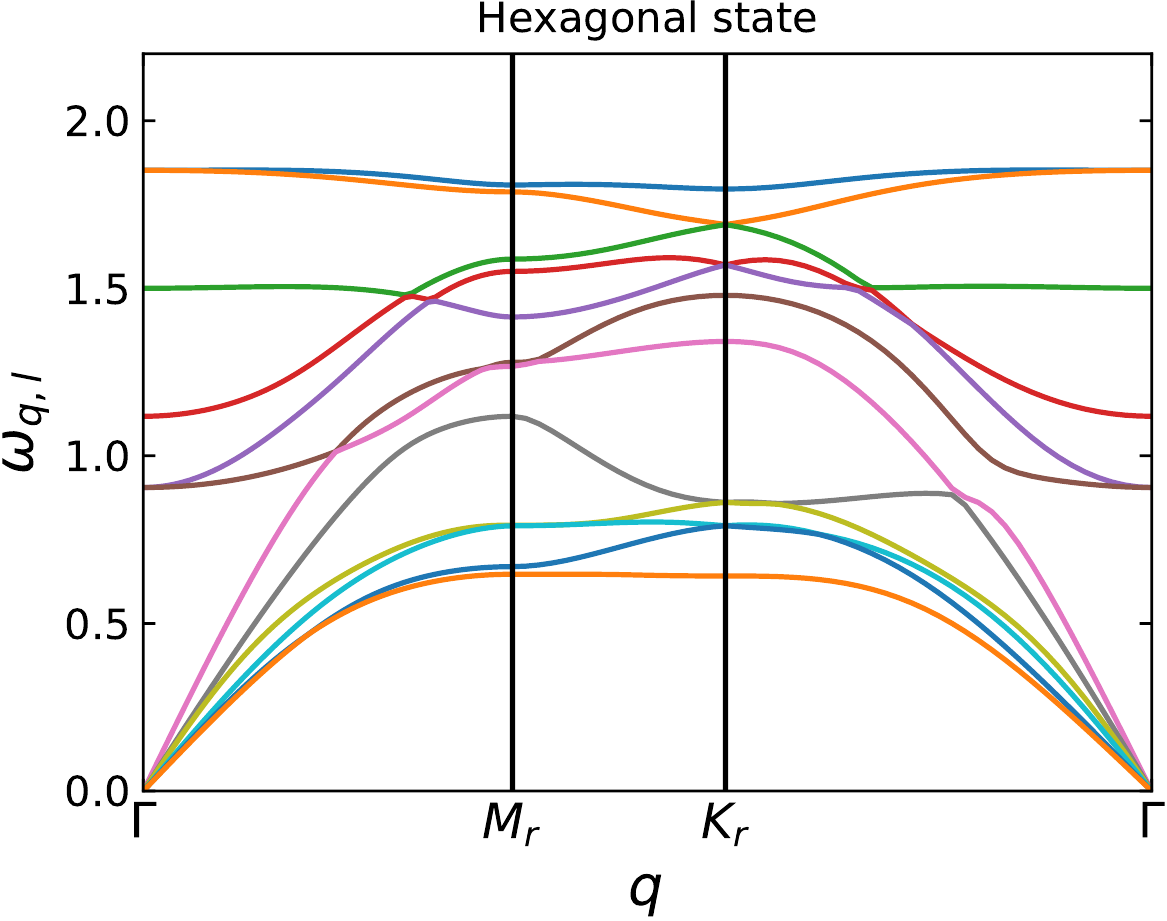}\quad
        \includegraphics[height=.22\textwidth, trim= 38 0 0 0, clip]{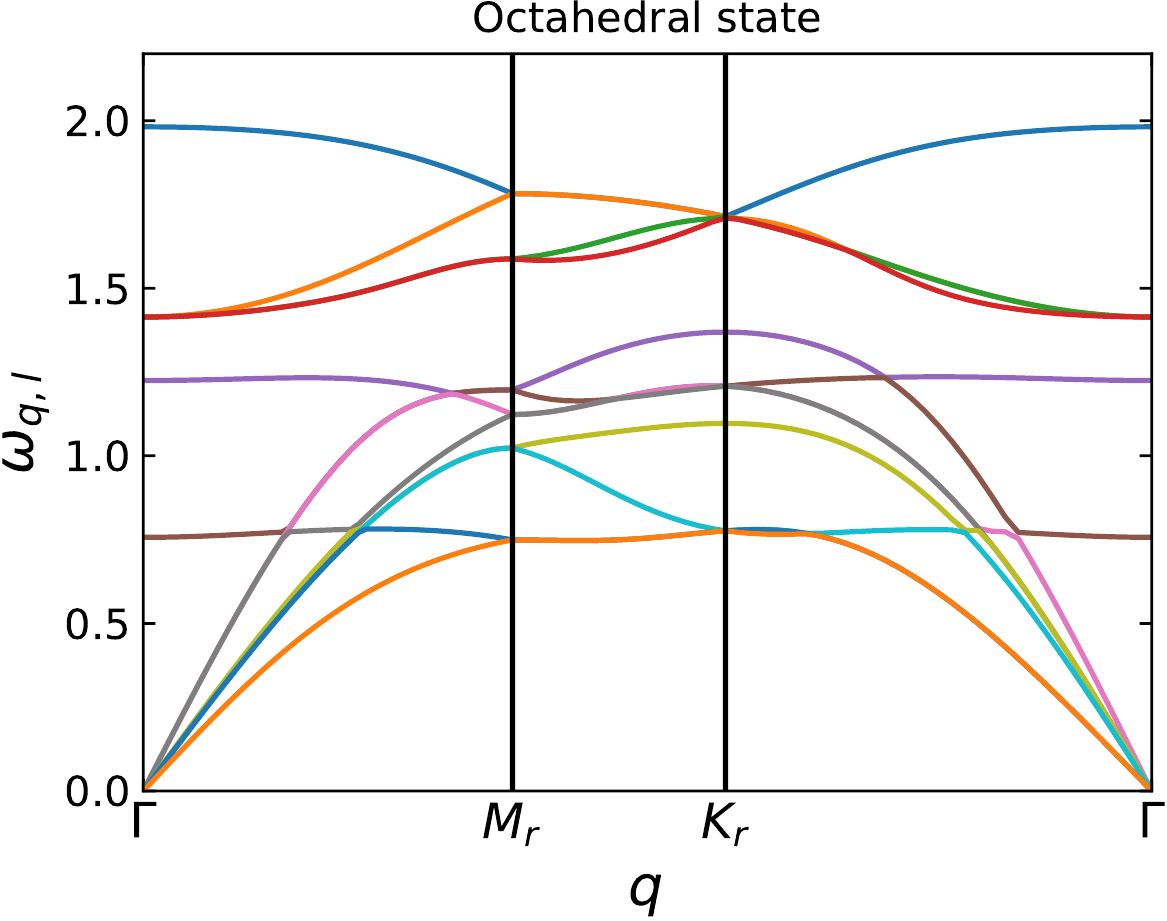}
        \quad
        \includegraphics[width=.14\textwidth, trim = 0 -90 0 0]{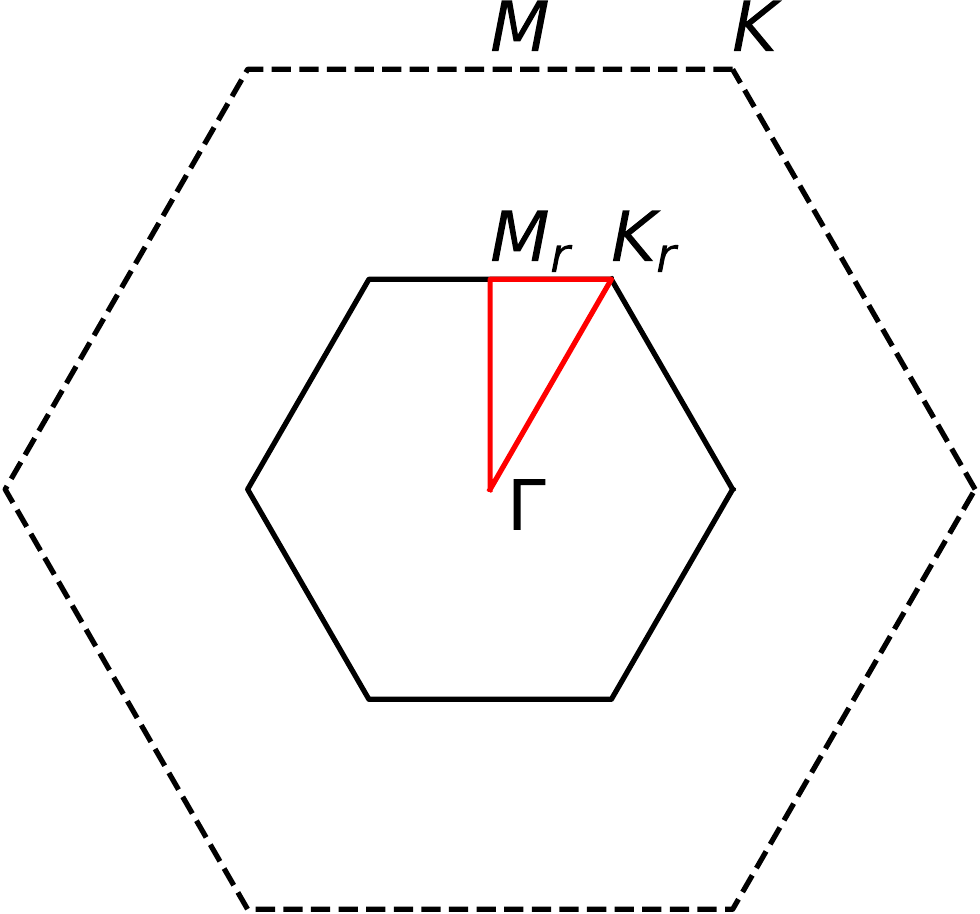}
        \caption{Dispersion relations $\omega_{\mathbf q, l}$ along a cut in the Brillouin zone (red line on the right) from linear spin wave approximation for $\phi = 3\pi/4$ ($J_3 = -J_1 >0$) for the $J_1-J_3$ kagom\'e model, for the three ground states of Fig.~\ref{fig:usual_orders_kagome}.
        As a unit cell of 12 sites has been chosen, there are 12 energy bands in the reduced Brillouin zone (full black line on the right). 
        }
        \label{fig:SW_dispersion} 
    \end{center}
\end{figure*}

The Hamiltonian \eqref{eq:Ham_SW} is now expanded in powers of $1/\sqrt S$. 
The first term is the energy of the reference classical ground state, in $S^2$. 
The next term, in $S^{3/2}$, is zero if the reference ground state has correctly been chosen, as a stationnary point of the reference energy with respect to the $R_i$'s. 
Finally, the first interesting term is in $S$, and has exactly the same form from Eq.~\eqref{eq:Uquant} or from \eqref{eq:Uclass}: it is a quadratic Hamiltonian either in $a_i$ and $a_i^\dag$ or in $z_i$ and $z_i^*$:
\begin{equation}
\label{eq:Ham_SW2}
H^S=
\frac 12\sum_{i,j} 
\mathbf v_i^\dag
M^S_{i,j}
\mathbf v_j
\end{equation}
where $M_{i,j}^S$ is a $2\times 2$ matrix and $\mathbf v_i$ is the two-component vector containing either $a_i$ and $a_i^\dag$ or $z_i$ and $z_i^*$. 

Depending on the periodicity of $M^S_{i,j}$, an eventually large unit-cell of $m$ sites is chosen to perform a Fourier transform $\mathbf {\tilde v}_{\mathbf q}$ of $\mathbf v_i$, of components:
\begin{equation}
\mathbf {\tilde v}_{\mathbf q} = 
\begin{pmatrix}
\tilde a_{\mathbf q, 1}\\
\tilde a_{\mathbf q, 2}\\
\dots\\
\tilde a_{\mathbf q, m}\\
(\tilde a_{-\mathbf q, 1})^\dag\\
(\tilde a_{-\mathbf q, 2})^\dag\\
\dots\\
(\tilde a_{-\mathbf q, m})^\dag
\end{pmatrix}, \quad
\begin{pmatrix}
z_{\mathbf q, 1}\\
z_{\mathbf q, 2}\\
\dots\\
z_{\mathbf q, m}\\
(z_{-\mathbf q, 1})^*\\
(z_{-\mathbf q, 2})^*\\
\dots\\
(z_{-\mathbf q, m})^*
\end{pmatrix}.
\end{equation}
The Hamiltonian rewrites:
\begin{equation}
\label{eq:Ham_SW3}
H^S=
\frac 12\sum_{\mathbf q} (\mathbf {\tilde v}_{\mathbf q})^\dag \cdot \tilde M^S_{\mathbf q} \mathbf {\tilde v}_{\mathbf q} + E_{\rm class}, 
\end{equation}
where $i$ and $j=1\dots m$ are now the indices of sites in the large unit cell and $\mathbf q$ are wave vectors of a reduced Brillouin zone. 
The constant $E_{\rm class}$ results from commutation relations used in the quantum case, and has no effect in the classical expansion. 

The eigenenergies $\omega_{\mathbf q, l}$ are determined via a Bogoliubov transformation, that preserves the bosonic commutation relations in the quantum case, and the conjugation relations between $z_i$ and $z_i^*$ in the classical case. 
We thus define new vectors $\mathbf {\tilde w}_{\mathbf q}$ from a matrix $P_{\mathbf q}$ such that $P_{\mathbf q}\mathbf {\tilde w}_{\mathbf q} = \mathbf {\tilde v}_{\mathbf q}$, with properties similar to the $\mathbf{\tilde v}_{\mathbf q}$, that are eigenmodes of the Hamiltonian (the transformed $\tilde M^S_{\mathbf q}$ matrix is diagonal). 
The information that we can extract from $P_{\mathbf q}$ and $\omega_{\mathbf q, l}$ in the quantum and classical cases will be described in the next subsections. 

We now apply this formalism to the $J_1-J_3$ model, in the octahedral part of the phase diagram (Fig.~\ref{fig:camembert}). 
A generic ground state is chosen, parametrized by three angles $\theta_B$, $\theta_C$ and $\phi_C$ where spins in the origin unit cell (on the green, blue and red sites of the marron triangle of Fig.~\ref{fig:Kag_3sublattices}) are:
\begin{equation}
\label{eq:SASBSC}
\mathbf S^{0}_A = \begin{pmatrix}0\\0\\1\end{pmatrix},\,
\mathbf S^{0}_B = \begin{pmatrix}\sin\theta_B\\0\\\cos\theta_B\end{pmatrix},\,
\mathbf S^{0}_C = \begin{pmatrix}\sin\theta_C\cos\phi_C\\\sin\theta_C\sin\phi_C\\\cos\theta_C\end{pmatrix}.
\end{equation}
This parametrization describes all the ground states, up to a global spin rotation (equivalent to an appropriate choice of the basis in the spin space). 
Moreover, up to a lattice translation, we can fix $0\leq \theta_B, \theta_C\leq \pi/2$, $0\leq \phi_C\leq \pi$. 
The three states of the bottom of Fig.~\ref{fig:usual_orders_kagome} are given from left to right by $(\theta_B, \theta_C, \phi_C)=(0,0,0)$ (colinear state), $(\pi/3,\pi/3,\pi)$ (hexagonal) and $(\pi/2, \pi/2, \pi/2)$ (octahedral). 

To perform the Fourier transformation of Eq.~\eqref{eq:Ham_SW3}, a unit-cell of 12 sites has to be chosen (as on Fig.~\ref{fig:Potts_order}), which results in $24\times24$ $\tilde M^S_{\mathbf q}$ matrices.
The dispersion relations for $\phi = 3\pi/4$ ($J_3 = -J_1 >0$) are given in Fig.~\ref{fig:SW_dispersion} for the colinear, hexagonal and octahedral states.

\subsection{Linear thermal spin wave approximation}
\label{sec:LTSW}

In two dimensions, we cannot expect to have a valid expansion at finite temperature: the Mermin-Wagner theorem predicts that a continuous order parameter (here the spin orientation), cannot survive to infinitesimal temperature. 
The hypothesis done on the small fluctuations around the classical ground state is false. 
However, short range correlations survive, and their nature can still be infered from entropic selection of the maximally fluctuating ground state at low temperatures\cite{Henley1987, Henley1989}. 

\begin{figure}
    \begin{center}
        \includegraphics[width=0.42\textwidth]{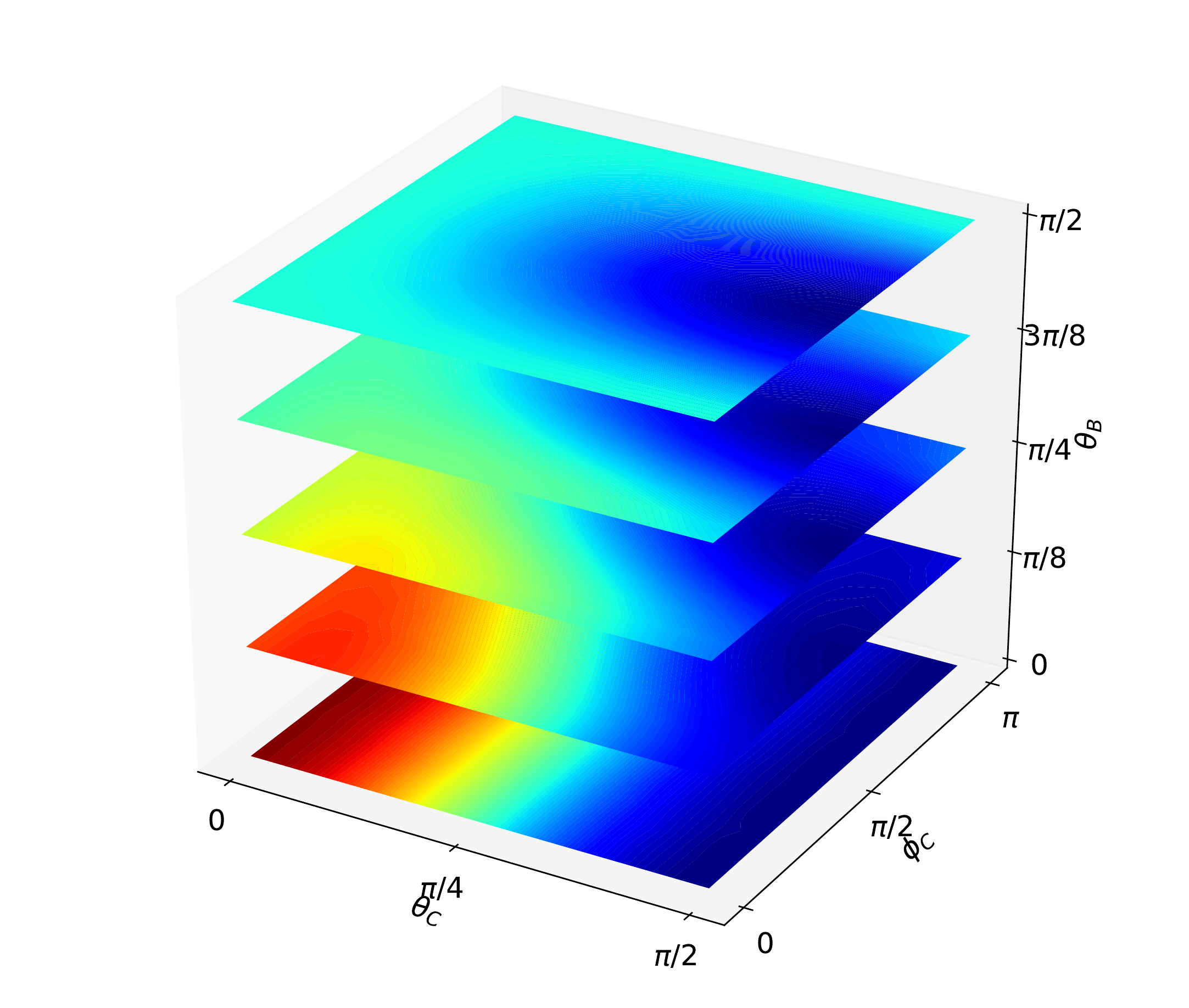}
        \includegraphics[width=0.42\textwidth]{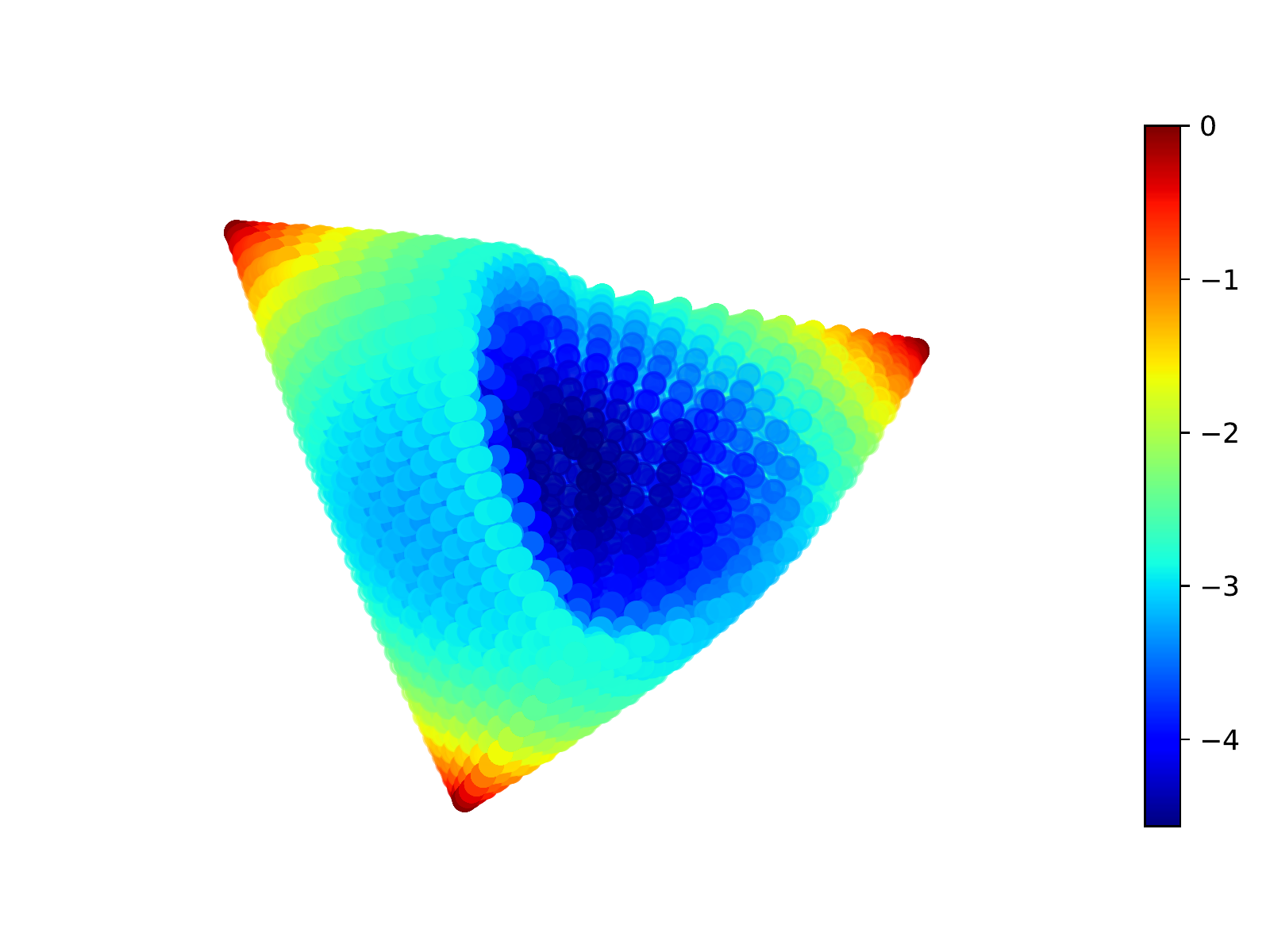}
        \caption{Low energy entropy $\Delta S(\theta_B,\theta_C, \phi_C)$ (top) and $\Delta S(\bm \sigma)$ (bottom), where the angles were defined in Eq.~\eqref{eq:SASBSC} and $\bm \sigma$ in Sec.~\ref{sec:order_parameter}, for $J_1<0$ and $J_3=-J_1$. 
        The maximal entropy (in dark red) is for $\theta_B=\theta_C=0$: the colinear state, and the minimum (in dark blue) for $\theta_B=\theta_C=\phi_C=\pi/2$: the octahedral state, corresponding respectively to the vertices and to the center of the inflated tetrahedron formed by the set of $\bm \sigma$ values. 
        }
        \label{fig:SW_S} 
    \end{center}
\end{figure}

Classical spins are described, in the linear spin wave approximation, by a collection of independent harmonic oscillators of frequencies $\omega_{\mathbf q, l}$. 
There are two modes for each couple $(\mathbf q, l)$, associated with the real ($x$ spin component) and imaginary ($y$ spin component) part of $z_i$ in Eq.~\eqref{eq:Uclass}. 
At finite temperature, the free energy $F=E-TS$ depends on the reference ground state which has been chosen. 
$E$ is the same for all of them, thus, it is the entropy that lifts the degeneracy. 
For a classical harmonic oscillator of frequency $\omega$, the entropy is $S= \rm{const} -\ln\frac{T}{\omega}$. 
A zero point energy is necessary to forbid negative values of the entropy at low temperature. 
The entropies of different reference ground states are parametrized by the angles $S(\theta_B, \theta_C,\phi_C)$ of Eq.~\eqref{eq:SASBSC}, or more conveniently, by the vector of spin dot-products $\bm \sigma$, defined in Sec.~\ref{sec:order_parameter}. 
The difference $\Delta S(\theta_B, \theta_C,\phi_C) = S(\theta_B, \theta_C,\phi_C)-S(0,0,0)$, or equivalently $\Delta S(\bm \sigma) = S(\bm \sigma)-S(\bm \sigma_0)$, where $\bm \sigma_0=(1,1,1)$, does not depend on the temperature and is represented on Fig.~\ref{fig:SW_S} for $\phi=3\pi/4$. 
The maximum is reached in the colinear state, and the minimum in the octahedral state, as expected. 

\subsection{Linear quantum spin wave approximation}
\label{sec:LQSW}

\begin{figure}
    \begin{center}
        \includegraphics[width=0.4\textwidth]{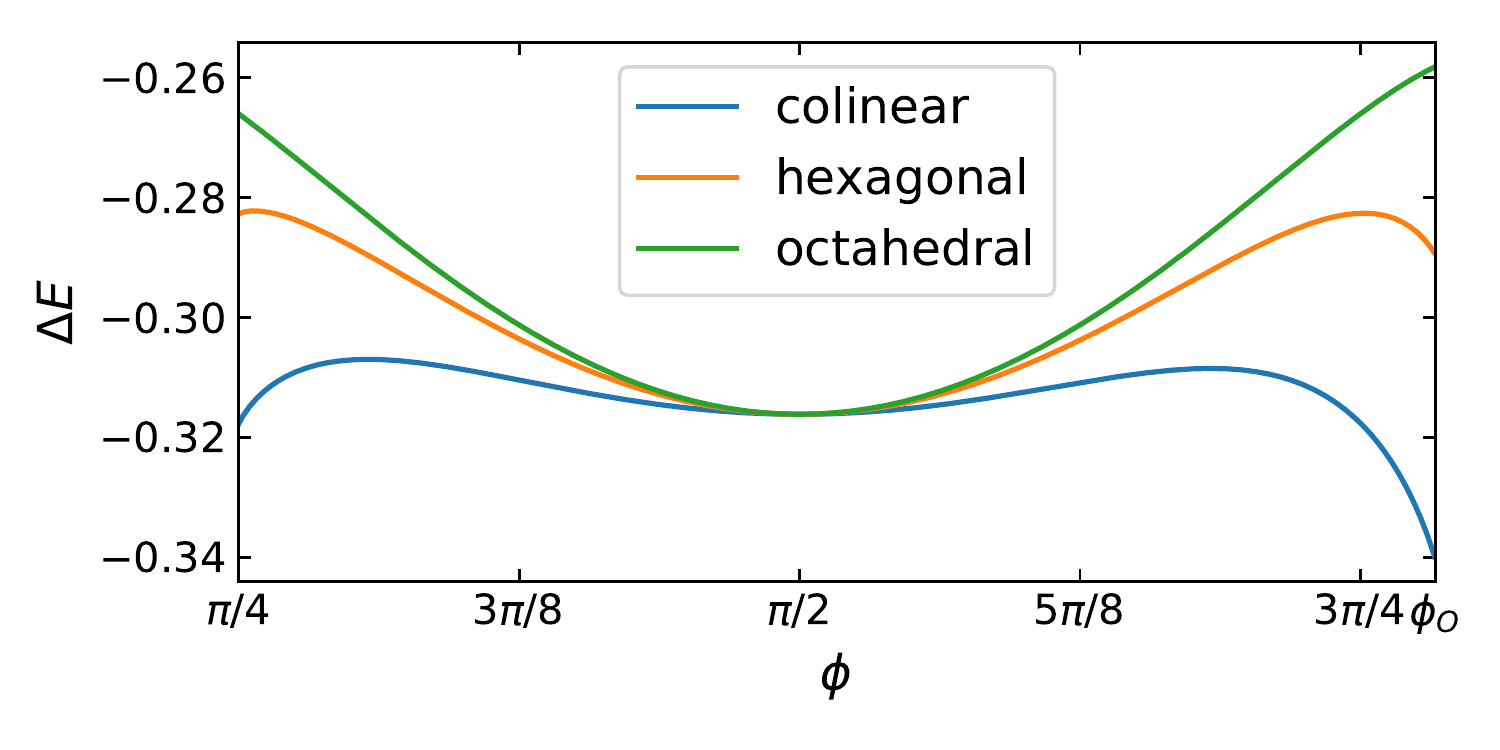}\\
        \includegraphics[width=0.4\textwidth]{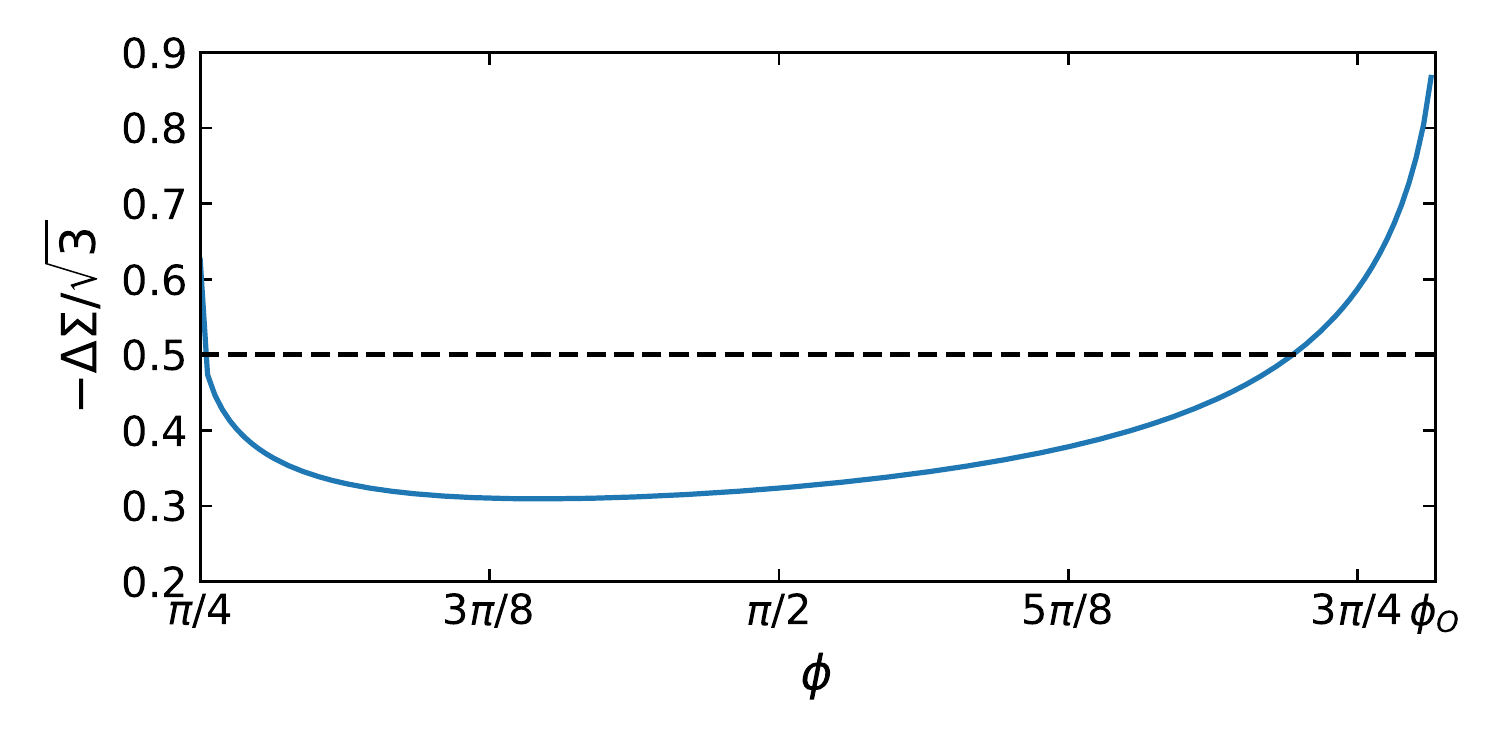}
        \caption{
            Results of quantum linear spin wave approximation for $(J_1, J_3)=(\cos\phi,\sin\phi)$, in the octahedral phase of Fig.~\ref{fig:camembert}. 
            Top: correction $\Delta E$ of order $S$ to the energy around the three classical states depicted in Fig.~\ref{fig:usual_orders_kagome}.
            Bottom: correction $\Delta \Sigma$ of order $S$ to the order parameter for the colinear phase. 
            The dashed line indicates an approximative value of $-\Delta\Sigma/\sqrt{3}$ above which quantum fluctuations restore the $K_4$ symmetry for $S=1/2$.
        }
        \label{fig:SW_DeltaE} 
    \end{center}
\end{figure}
 
In quantum materials, the spin has a finite value ($S=1/2$, $1$, $3/2$...), which differs from the classical case corresponding to the limit $S\to \infty$. 
In Ba-Vesignieite, the spin on the copper sites has the most quantum value of $1/2$. 
We now discuss the consequences in light of the previous classical considerations. 
Quantum fluctuations tend to disorder the system: a model with a magnetically ordered ground state in the classical limit generally has an order parameter $m$ that decreases when $S$ decreases. 
We thus face two possibilities: either the order parameter remains finite ($m>0$) when quantum fluctuations are switched on, or it reaches zero and the ground state is no more long-range ordered. 
 
The linear spin wave approximation expands to first non trivial order quantum observables (as the energy or an order parameter) in $1/\sqrt{S}$ at zero temperature and around a specific ground state. 
When several ground states exist, as occurs here in the $J_1-J_3$ model, the expansion can be performed around any of them, giving different correction to the energy that eventually lifts the degeneracy. 
The first terms of the energy are:
\begin{equation}
E = S(S+1) E_{\rm class} - \frac S2 \sum_{\mathbf q, l} \omega_{\mathbf q, l} + \mathcal O(\sqrt{S}),
\end{equation} 
where $E_{\rm class} = -2J_3$ in the octahedral phase. 
The term of order $S$: $\Delta E = E_{\rm class}-\frac 12\sum_{\mathbf q, l} \omega_{\mathbf q, l}$, depends on the angles $(\theta_B, \theta_C, \phi_C)$ and on the coupling $\phi$. 
It can be represented in the same way as $\Delta S$ in Fig.~\ref{fig:SW_S} for a fixed $\phi$. 
The same qualitative behavior is obtained, and the same conclusion: the colinear state is the most favored by quantum fluctuations, whereas the octahedral one has the weakest quantum energy correction. 
It is quite expected that quantum and thermal fluctuations favor the same order, even if counter-examples exist\cite{PhysRevLett.105.265301}. 
For completeness, the curve of $\Delta E$ is given versus $\phi$ in Fig.~\ref{fig:SW_DeltaE}, for the three ground states of Fig.~\ref{fig:usual_orders_kagome}. 
Whatever $\phi$ (except $\phi= \pi/2$ where the three sublattices are completely decoupled), quantum fluctuations always favor the colinear state. 

The order parameter $\Sigma$ can be expanded as the energy: $\Sigma = S^2\,\Sigma_{\rm class} + S\Delta \Sigma + O(\sqrt S)$, which can be used as an indication of the critical spin where its average cancels, excluding the occurence of a phase transition as finite temperature. 
The classical value is $\Sigma_{\rm class} = \sqrt{3}$. 
Thus, $S_c\sim -\frac{\Delta \Sigma}{\sqrt 3}$.
$S_c$ is below $1/2$ in all the octahedral phase, except near the boundary with the unconventional phase (Fig.~\ref{fig:SW_DeltaE}). 
It suggesting that the $K_4$ symmetry could be broken even in the $S=1/2$ case.


\section{High temperature series expansions (HTSE)}
\label{sec:HTSE}

\begin{figure}
	\begin{center}
		\includegraphics[width=0.4\textwidth]{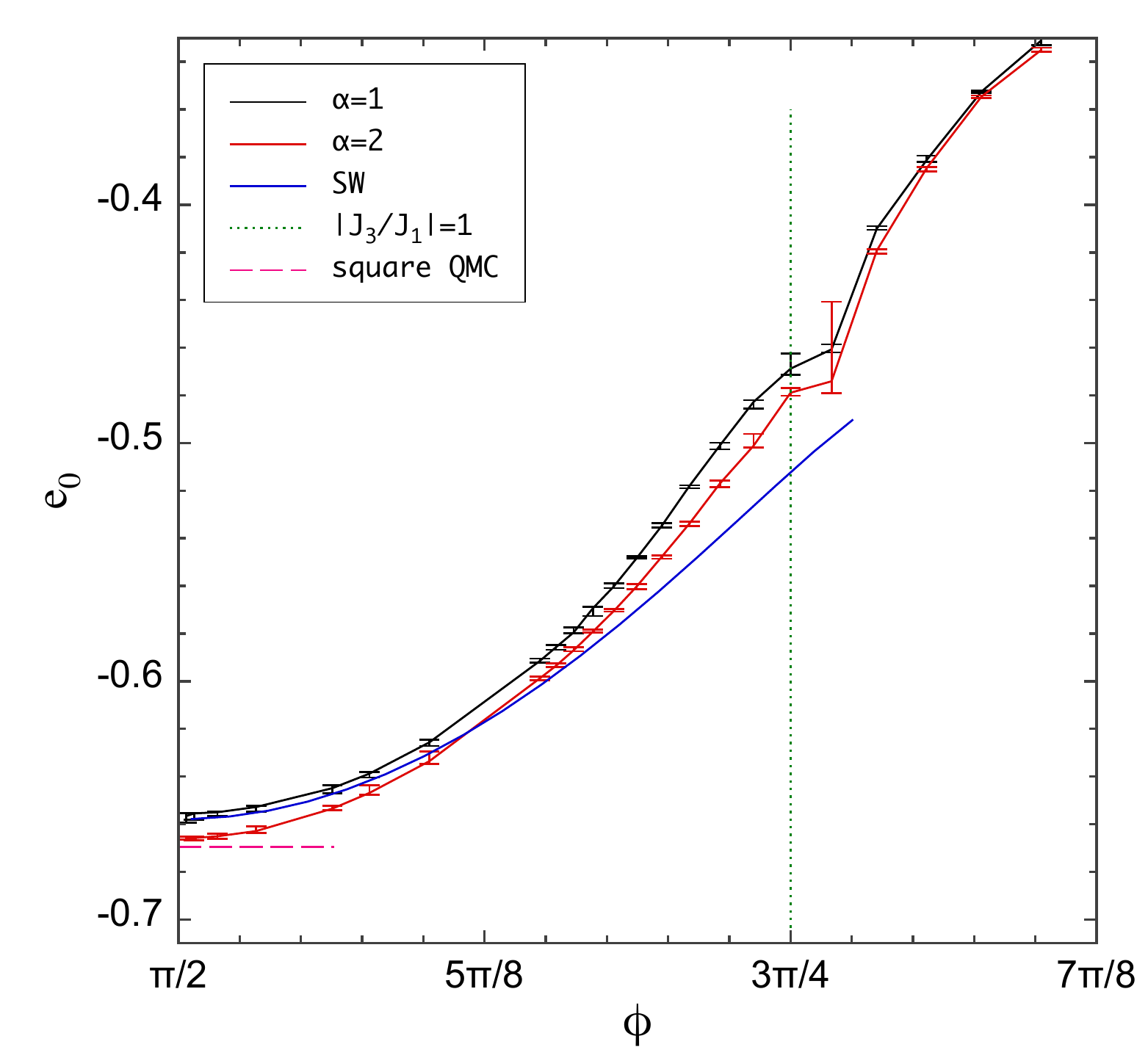}
		\caption{Ground state energy per site $e_0$ as a function of $\phi$, with $(J_1,J_3)=(\cos\phi, \sin\phi)$ on the kagom\'e lattice. 
        $e_0$ is obtained via the method described in \textcite{Bernu2020}, using high temperature series expansions up to order 15. 
		The red and black points are the results with the hypothesis that $C_V \sim A T^\alpha$, with $\alpha=1$ and $2$ and $A$ a constant. 
		The blue curve is the linear spin wave energy up to order $S$, approximated for $S=1/2$.
		}
		\label{fig:HTSE} 
	\end{center}
\end{figure}

After a look at the behavior of the model from the classical limit ($S=\infty$) towards finite spins, the extreme quantum case of $S=1/2$ can be investigated through high temperature series expansions. 
The logarithm of the partition function $\frac{\ln Z}N(\beta)$ is expanded in powers of the inverse temperature $\beta$ directly in the thermodynamic limit:
\begin{equation}
\lim_{N\to\infty} \frac{\ln Z}N(\beta) = \ln 2 + \sum_{n=1}^\infty \left( \sum_{i=0}^n Q_{i,n} J_1^i J_3^{n-i}\right) \beta ^n , 
\end{equation}
where $N$ is the number of lattice sites. 
Enumerating connected clusters on the $J_1-J_3$ kagom\'e lattice, we exactly calculate the coefficients of this series up to order 15 in $\beta$, each of them being an homogeneous polynom in $J_1$ and $J_3$. 

A direct use of the truncated series to evaluate thermodynamical functions is doomed to fail, as the series only converges for $T \gtrsim J_1, J_3$. 
An extrapolation technique called the entropy method (HTSE$+s(e)$) has been developpedd\cite{PhysRevB.63.134409, Bernu2015}, that extrapolates functions from infinite down to zero temperatures. 
It uses the hypothesis of the absence of finite temperature phase transition, so that the functions are analytical over the full temperature interval. 
It also requires some inputs: the ground state energy per site $e_0$ and the low temperature behavior of $C_V$ (in power law $C_V\sim T^\alpha$, or exponential for example), what can be understood as the need to constrain the thermodynamical functions both from the $T=\infty$ side, which is ensured by the series coefficients, and from the $T=0$ one.

The need for $e_0$ is a real problem, as no generic method exist to determine it in the case of frustrated quantum models. 
In \textcite{Bernu2020}, a self-consistent method has been developed that proposes an $e_0$. 
Although no rigorous argument says that this energy is near the real one, it has been shown to give extremely coherent results on the first neighbor kagome model. 
With the hypothesis that no phase transition occurs, the ground state energy $e_0$ obtained by this method is shown in Fig.~\ref{fig:HTSE}, for $C_V\sim_{T\to0}A\,T^\alpha$ with $\alpha=2$ (which is the case for $\phi=\pi/2$) and $\alpha=1$. 
The minimal $\phi=\pi/2$ on Fig.~\ref{fig:HTSE} corresponds to the three decoupled square sub-lattices, whose ground state energy is accessible through quantum Monte Carlo simulations in this unfrustrated case: $e_0=-0.6695$\cite{PhysRevLett.80.2705, PhysRevB.57.11446}. 
HTSE$+s(e)$ results give still better results that the linear spin wave approximation. 
With increasing $\phi$, error bars increase and the result quality becomes bad in the neighborhood of $\phi_O$ (convergence issue of the method), at the point where a slope breaking occurs in $e_0(\phi)$. 

In view of the previous sections, this behavior can be attributed to the existence of a phase transition at finite temperature $T_c$ near $\phi_O$. 
In the Supp. Mat. of \cite{Bernu2015}, the possibility to detect a phase transition thanks to HTSE$+s(e)$ was proposed for a ferromagnetic BCC lattice, where $e_0$ was exactly known and the extrapolation was performed down to $T=0$ despite the singularity at $T_c$. 
Here, the method tends to deviate $e_0$ from its real value to get ride of eventual singularities. 
We propose a new adaptation of HTSE$+s(e)$ to models with phase transitions, that will be detailed elsewhere\cite{HTE_transition}. 
The extrapolation is only done on the temperature interval $\lbrack T_c,\infty\rbrack$, requiring as supplementary input parameters $T_c$, the energy $e_c$ and the entropy $s_c$ at $T_c$
We also characterize the behavior of $C_V$ near the transition by an exponent $\alpha$:
\begin{equation}
C_V(T) \sim_{T\to T_c^+} \frac{A}{(T-T_c)^\alpha}. 
\end{equation}
Because of the sum rules on $C_V(T)/T$, $\alpha$ must be lower or equal to 1. 
For $J_1=-1$ and $J_3=1$, the four parameters $T_c$, $e_c$, $s_c$ and $\alpha$ giving the higher quality of result were looked for. 
Interesting values are found in a tiny valley of the 4 dimensional space, with a transition at $T_c=0.42(1)$ and an exponent of $\alpha = 0.29(1)$, $e_c=-0.405(5)$ and $s_c=0.35(1)$. 

Even if still exploratory, this section on HTSE confirm the possibility of a phase transition in the $S=1/2$ model, in the domain of parameter where it is the more easily detected in the classical model: $J_3\simeq |J_1|$.

\section{Conclusion}
\label{sec:conclusion}

Motivated by the Ba-Vesignieite compound, this article has explored the $J_1-J_3$ model on the kagom\'e lattice, in the domain of large $J_3$. 
The classical phase diagram has revealed interesting phases: for ferromagnetic $J_1$ and moderate $J_3$, an unconventional phase displays conical, spiral, and probably other unusual phases, whereas for large $J_3$, whatever the sign of $J_1$, an octahedral phase possesses an accidental degeneracy. 
Thermal or quantum fluctuations lift this degeneracy via the order by disorder mechanism, favouring colinear configurations, labelled by an element of the $K_4$ group. 
An order parameter $\bm \Sigma$ was constructed by analysing the symmetries of the model, to detect this discrete $K_4$ symmetry breaking. 

Classical Monte Carlo simulations have evidenced an order-by-disorder induced phase transition associated with $\Sigma$. 
The transition is first order for low $J_3$'s, and either weakly first order or second order for large ones. 
Other phase transitions were found in the unconventional phase, associated with one or several other order parameters. 

Linear spin wave formalism have shown that both thermal and quantum fluctuations favor the colinear states. 
But quantum fluctuations can be so strong that they completely disorder the system, preventing the occurence of a phase transition, notably near the boundary with the unconventional phase $\phi=\phi_O$. 
Finally, HTSEs also confirm the possibilitiy of a phase transition, this time in the $S=1/2$ model. 

What are the implication of this phase transition on Ba-Vesignieite ? 
First of all, the dominant coupling was proposed to be $J_3$ in \cite{PhysRevLett.121.107203}, but the one coming next was $J_3'$, then $J_1$, and $J_2$. 
We did not considered $J_3'$ as it did not couple the three kagom\'e sublattices, and have focused on $J_1$. 
Note that $J_2$ would have led to the same order by disorder effect as $J_1$. 
One could argue that many perturbations other than next nearest neighbor interactions can lift the degeneracy of the octahedral phase. 
Among them, a slight distortion of the lattice is know, of less that 1\% of the Cu-Cu distance and causes a coupling anisotropy\cite{PhysRevB.83.180416}. 
Some impurities are unavoidable, whose effect has been studied on the $J_1-J_2$ square lattice. Their effect is opposite to the one of thermal fluctuation, selecting orthogonal configurations\cite{Henley1987, Henley1989}, and penalizing colinear ones. 
If this occurs here, the octahedral state of Fig.~\ref{fig:usual_orders_kagome} would be favored, possibly leading to a chiral phase transition. 
\DM interactions must also be present\cite{Zorko2013}, as well as Ising spin anisotropy\cite{PhysRevLett.121.107203} but eventually very small. 
Lastly, a small coupling between spins in successive kagom\'e planes exists and is suspected to induce the phase transition observed at $T=9K$\cite{PhysRevLett.121.107203}. 

However, despite this whole set of deviations from the $J_1-J_3$ model, the transition discussed in this article remains meaningful. 
At temperature larger than their typical value, their effect is crushed, and the $K_4$ order can still be present. 

Lastly, the theorical investigation of such an emerging $q=4$ Potts order parameter and of its phase transition illustrate in an original way the order by disorder mechanism. 


\section*{Acknowledgments}
We thank Bjorn F\aa k for discussions on the experimental results on Vesignieite.
This work was supported by the French Agence Nationale de la Recherche under Grants No. ANR-18-CE30-0022-04 LINK. 

Numerical simulations were performed on the highly parallel computer of the LJP, LKB and LPTMC.

\appendix

%

\section{The Luttinger-Tizsa method}
\label{app:LT}

To use the LT method, we perform a Fourier transform on $H$. 
With this in mind, we rewrite Eq.~\eqref{eq:HamJ1J3}
\begin{equation}
H = \frac{1}{2}\sum_{\mathbf{r}}\sum_{\mathbf{v}}\sum_{i, j} J_{i,j}(\mathbf{v}) \, {\mathbf S}_{i, \mathbf{r}} \cdot {\mathbf S}_{j, \mathbf{r}+\mathbf{v}},
\label{eq:ham_complet}
\end{equation}
where $\mathbf{r}$, $\mathbf{r+v}$ are vectors from a Bravais lattice locating the unit-cells of the interacting spins, $i$ and $j$ label inequivalent sites in each unit-cell. 
Next, we introduce the Fourier modes of a spin $i$ in cell $\mathbf{r}$:
\begin{equation}
\mathbf{S}_{i,\mathbf{r}} = \frac{1}{\sqrt{N}} \sum_{\mathbf{q}} \textbf{\~{S}}_i(\mathbf{q})e^{i\mathbf{q}\cdot\mathbf{r}}, 
\end{equation}
to rewrite the Hamiltonian as:
\begin{equation}
\label{eq:HamJijQ}
H = \frac{1}{2}\sum_{\mathbf{q}}\sum_{i, j} \textbf{\~{S}}_i(\mathbf{q}) \tilde J_{i,j}(\mathbf{q})  \textbf{\~{S}}_j(\mathbf{-q})
\end{equation}
where $\tilde J_{i,j}(\mathbf{q}) = \sum_{\mathbf{v}} J_{i,j}(\mathbf{v}) e^{i\mathbf{q}\cdot\mathbf{v}}$ is akin to a Fourier transform of the couplings of $H$. 
The Hamiltonian itself is now expressed as a bilinear form in the Fourier modes $ \textbf{\~{S}}_i(\mathbf{q})$. Its ground state may easily be found by diagonalizing $\tilde J(\mathbf{q})$ and minimizing its lowest eigenvalue $\lambda_{\rm min}(\mathbf{q})$ with respect to $\mathbf{q}$. 
This, in turn, leads us to a generally discrete set of wave vectors $\mathbf{q}_i$ of the Brillouin zone respecting the lattice symmetries\cite{Villain1977}. 
The desired ground state is then obtained by solely populating the eigenmodes corresponding to $\lambda_{\rm min}(\mathbf{q}_i)$ and performing an inverse Fourier transform.
\\
In the preceding paragraph we never mentioned the nature of the lattice - Bravais or not - 
in order to justify the steps we took. 
One can wonder why, then, is it not possible to apply the LT methodology to our particular instance of the problem. 
The answer lies in an unmentionned constraint we ought to abide by: at each site we have a unit spin $\mathbf{S}_i$, with ${\|\mathbf{S}_i\|}=1$. 
For Bravais lattices this is not an issue, since there always exists a spiral state, defined by a single 
wavevector, which is a ground state of the Hamiltonian. 
For non-Bravais lattices, however, such as the kagom\'e lattice we're working on, this constraint prevents 
us from applying the last step, as naively populating a mode with the lowest energy generally does not respect the constraint on all sites of a unit cell.
Thus, other modes can be used to recover the constraint, increasing the energy as compared with $\lambda_{\rm min}$, which is then only a lower bound.

\section{Summary of results on the Potts model}
\label{App:exponents_q4}

The critical exponents of the two-dimensional Potts mode have a conjectured exact expression for $q\leq 4$ (See \cite{Wu1982}), that leads to:
%
\begin{equation}
\begin{tabular}{ccc}
$\alpha=\frac{2}{3}$,&
$\gamma=\frac{7}{6}$,&
$\beta =\frac{1}{12}$,
\nonumber\\
$\delta=15$,&
$\nu=\frac{2}{3}$,&
$\eta=\frac{1}{4}$
\end{tabular} 
\end{equation}

\section{Determination of the value of \texorpdfstring{$\phi_O$}{phiO}}
\label{app:phiO}

\begin{figure}[t!]
	\begin{center}
		\includegraphics[width=0.23\textwidth]{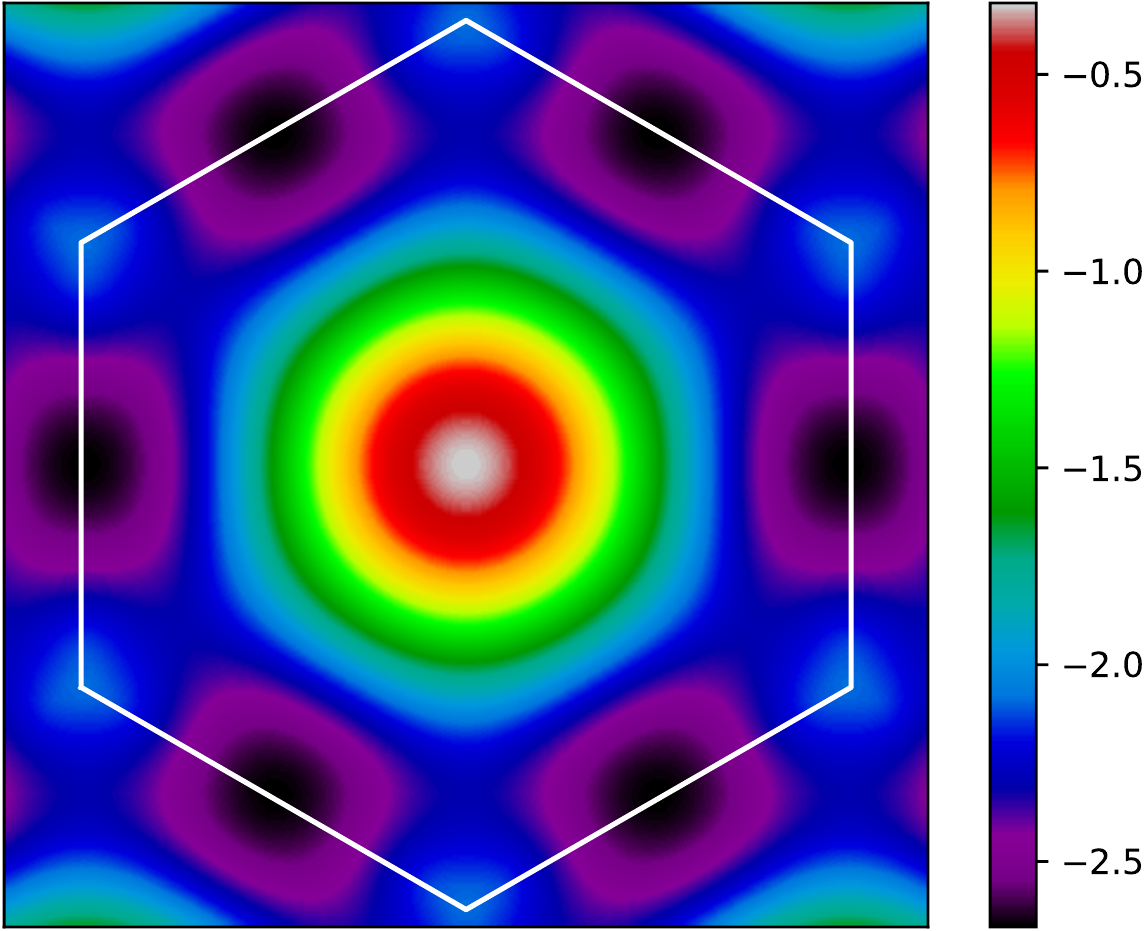}
		\includegraphics[width=0.23\textwidth]{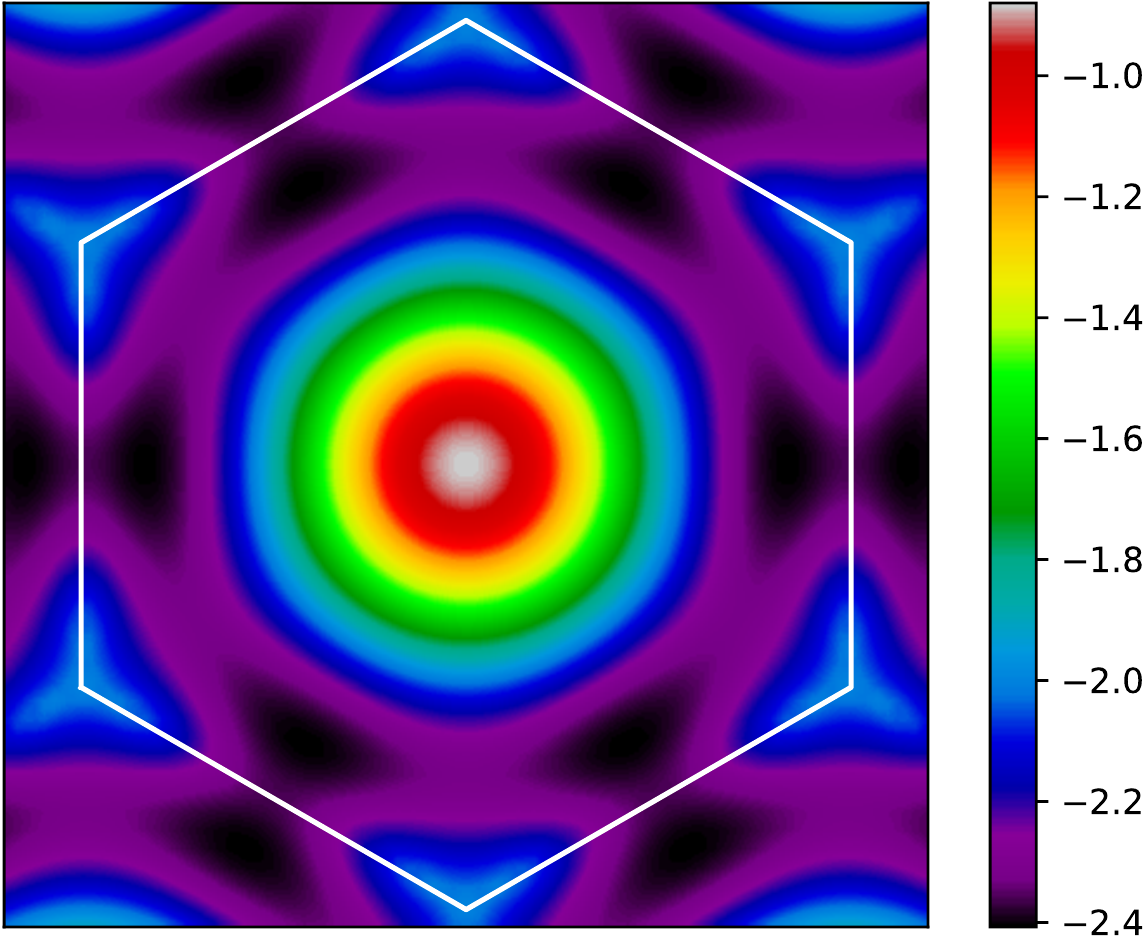}\\
		\includegraphics[width=0.23\textwidth]{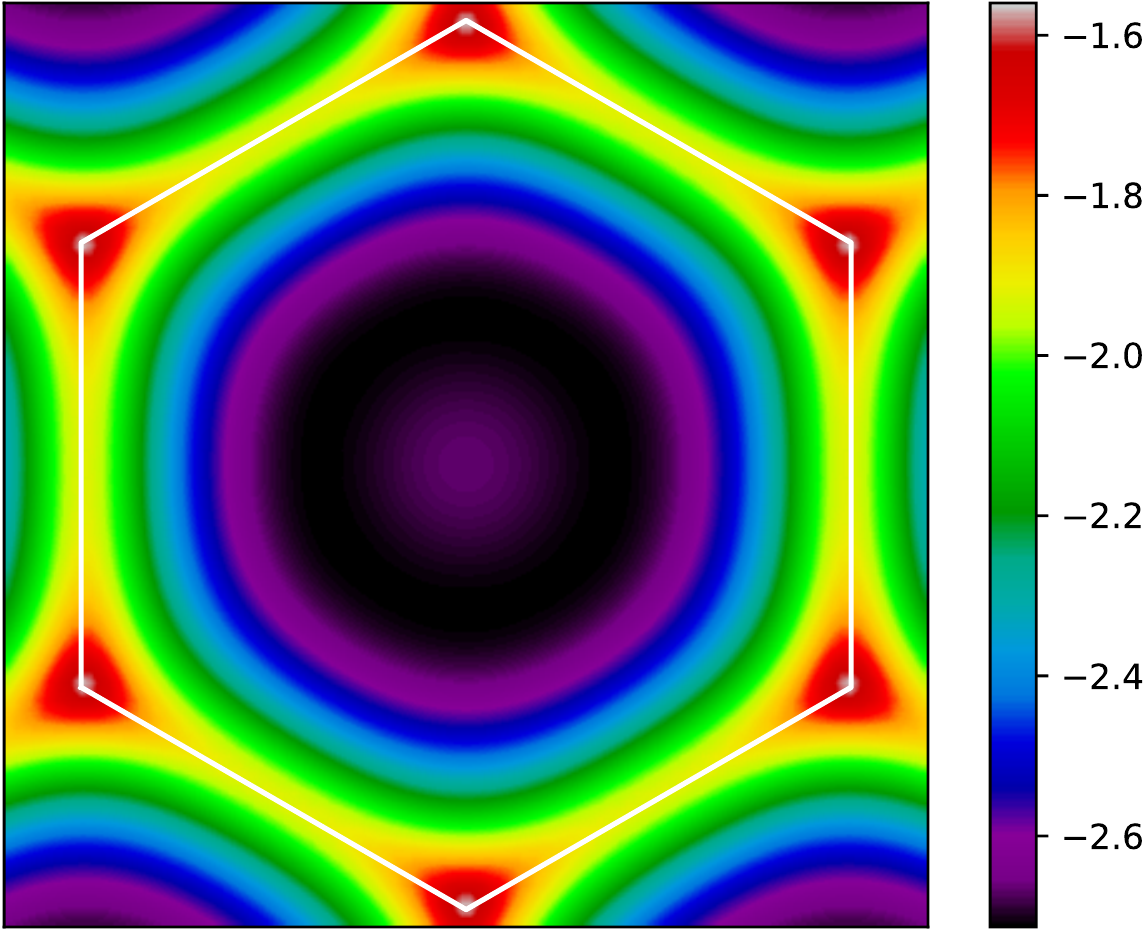}
		\includegraphics[width=0.23\textwidth]{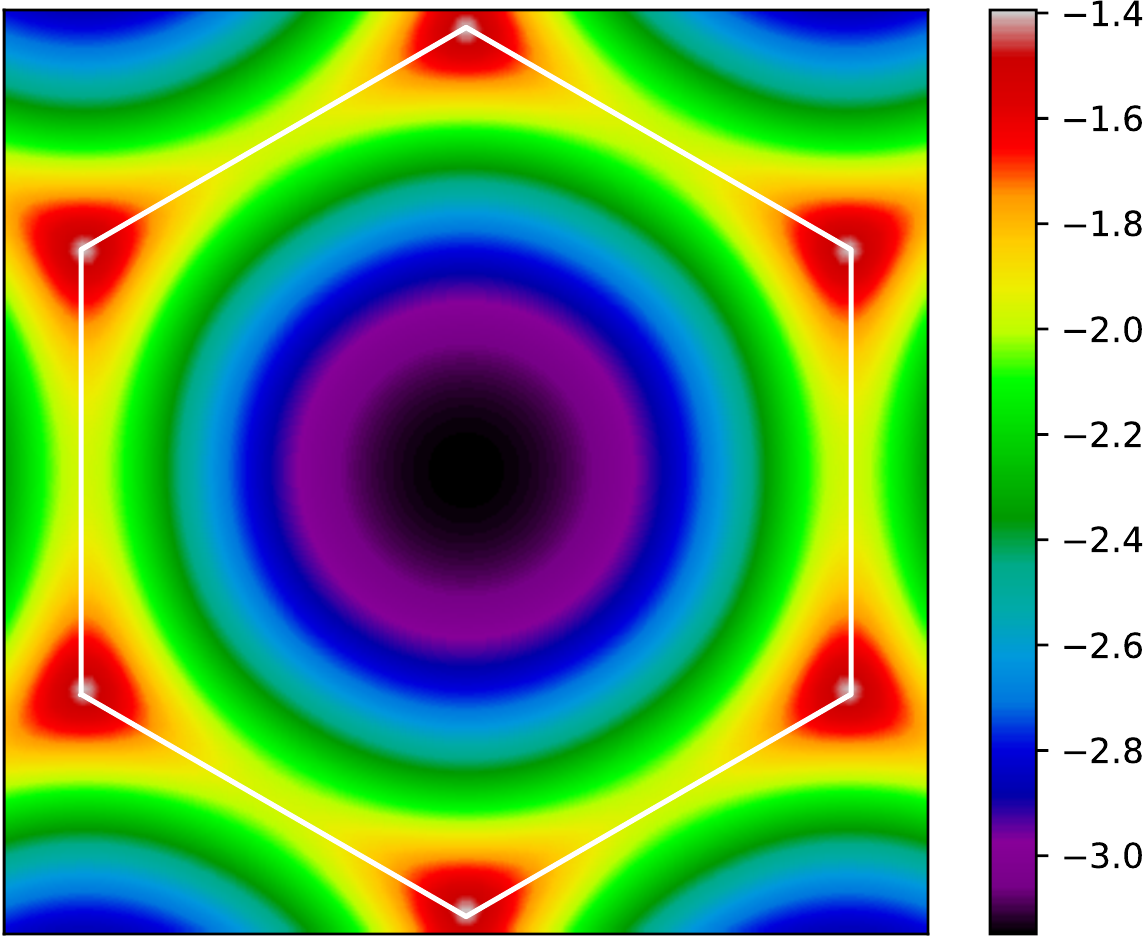}
		\caption{$\lambda_{\rm min}(\mathbf{q})$ for the $J_1-J_3$ model 
			on the kagom\'e lattice with $\phi=\phi_O-0.05$ (top left), $\phi_O+0.05$ 
			(top right) $\phi=\phi_F-0.05$ (bottom left), $\phi_F+0.05$ (bottom right). }
		\label{fig:LT_J1J_3_kag} 
	\end{center}
\end{figure}

We present here a derivation of the value of $\phi_O$, where the transition between the orthogonal and unconventional phase occurs in the $J_1-J_3$ model on the kagom\'e lattice (see Fig.~\ref{fig:camembert}). 
The proof rests on the LT method presented in App.~\ref{app:LT}. 
The $\tilde J(\mathbf{q})$ matrix of Eq.~\eqref{eq:HamJijQ}, multiplied by the overall $\frac{1}{2}$, writes:
\begin{equation}
\begin{pmatrix}
J_3 (c_1^2 + c_2^2 - 1)
&
\frac{J_1}2 c_1
&
\frac{J_1}2 c_2
\\
\frac{J_1}2 c_1
&
J_3(c_1^2 + c_3^2-1)
&
\frac{J_1}2c_3
\\
\frac{J_1}2c_2
&
\frac{J_1}2 c_3
&
J_3(c_2^2 + c_3^2-1)
)
\end{pmatrix},
\end{equation}
where $c_1 = \cos \frac{q_x}2$, $c_2 = \cos \frac{q_y}2$ and $c_3 = \cos \frac{q_x-q_y}2$.

In the octahedral phase, the minimal eigenvalue $\lambda_{\rm min}(\mathbf{q})$ of $\tilde J(\mathbf{q})$ occurs for three $\mathbf q$: $\mathbf M_{1,2,3}$, the middles of the edges of the Brillouin zone (see Fig.\ref{fig:LT_J1J_3_kag}). 
At the transition toward the unconventional phase, each of the three minima splits in two, giving six new minima evolving with $\phi$ along the line $M_i-\Gamma$.

The characteristic polynomial $C(\lambda)$ of the $\tilde J_{i,j}(\mathbf{M}_1+\mathbf{\delta_q})$ matrix is expanded to the first order in $\epsilon = \lambda + 2 J_3$, as we look for the minimal root of $C(\lambda)$, which is nearby $-2 J_3$ (the energy of an octahedral state) in the neighborhood of $\mathbf M_1=(\pi,0)$ and for the values of $J_3$ and $J_1$ of interest. 
The root of the first order degree polynomial approximating $C(\lambda)$ is expanded to the second order in $\mathbf {\delta q}$. 
Increasing from $\phi=\pi/2$, the quadratic form thus obtained changes at $\phi=\phi_t=\pi-\arctan \frac{1+\sqrt{5}}{4}$ from a positive one, with a minima at $\mathbf{\delta q=0}$, to a non-positive one, with a saddle point at $\mathbf{\delta q=0}$, indicating that the energy of the octahedral state is no more the lower bound, and that $\phi_O\geq \phi_t$ (they are unequal if the octahedral phase remains the ground state in the area where it does not have the LT lower bound energy). 

It remains to exhibit a state that has a lower energy than the octahedral state for $\phi>\phi_t$ to prove that $\phi_t$ is effectively the transition value. 
This is done using the conical state of Fig.~\ref{fig:config_unconventional}. 
We parametrize it by four angles $(\theta, \phi, \psi)$. 
A unit cell of 12 sites is defined as indicated on Fig.~\ref{fig:config_unconventional}, with three different spin orientations $\mathbf S_{1,2,3}$. 
A translation in the $\mathbf e_1$ direction has no effect on the spin orientation, whereas a translation in the $\mathbf e_2$ ($y$ coordinate) rotates the spins of $\phi$ and inverse them:
\begin{eqnarray}
\mathbf S_1 &=& (-1)^y \begin{pmatrix}
\cos 2y\psi \\ \sin 2y\psi \\ 0
\end{pmatrix},
\nonumber\\
\mathbf S_2 &=& (-1)^y \begin{pmatrix}
-\sin\phi\sin((2y-1)\psi) \\ \sin\phi\cos ((2y-1)\psi) \\ \cos\phi
\end{pmatrix},
\nonumber\\
\mathbf S_3 &=& (-1)^{y} \begin{pmatrix}
-\sin\phi\sin((2y-1)\psi) \\ \sin\phi\cos ((2y-1)\psi) \\ -\cos\phi
\end{pmatrix}
\end{eqnarray}
The energy per site thus reads:
\begin{eqnarray}
E&=&
\frac{2J_1\sin\phi}3
(\sin\psi(1-\cos 2\psi) -\cos\psi \sin 2\psi  + \sin\phi )
\nonumber\\
&&
+\frac{2J_3}3(
2\sin^2\phi\sin^2\psi - \cos 2\psi - 2\cos^2\phi)
\nonumber
\end{eqnarray}
The minimum of this energy (numerically obtained) is effectively between the lowest bound and the energy of the octahedral state for $\phi\gtrsim\phi_O$ (see the inset of Fig.~\ref{fig:camembert}, bottom)

\bibliography{prb2}

\end{document}